\documentclass[12pt]{article}
\usepackage{graphicx}
\usepackage{subfigure}
\usepackage{latexsym}
\usepackage{amsmath}
\usepackage[latin1]{inputenc}
\newcommand{\beqs}{\begin{equation*}}

\newcommand{\eeqs}{\end{equation*}}

\newcommand{\beqas}{\begin{eqnarray*}}
\newcommand{\beqa}{\begin{eqnarray}}

\newcommand{\eeqas}{\end{eqnarray*}}
\newcommand{\eeqa}{\end{eqnarray}}

\newcommand{\blist}{\begin{itemize}}

\newcommand{\elist}{\end{itemize}}

\providecommand{\href}[2]{#2}

\DeclareFontFamily{OT1}{rsfs}{}
\DeclareFontShape{OT1}{rsfs}{m}{n}{ <-7> rsfs5 <7-10> rsfs7 <10->rsfs10}{} 
\DeclareMathAlphabet{\mycal}{OT1}{rsfs}{m}{n}

\def\cM{{\cal M}}

\def\extd{{\rm d}}

\newcommand{\pd}[2]{\frac{\partial {#1}}{\partial {#2}}}

\def\tr{{\rm tr}}
\def\Tr{{\rm Tr}}
\def\TrL2{{\rm Tr}_{L^2}}

\def\atbdry{\Big|_{\partial \cM}}
\def\atbdry0{\Big|_{\partial \cM_0}}
\def\atbdry1{\Big|_{\partial \cM_1}}

\newcommand{\beq}{\begin{equation}}
\newcommand{\eeq}{\end{equation}}
\newcommand{\bea}{\begin{eqnarray}}
\newcommand{\eea}{\end{eqnarray}}

\begin{document}
\begin{titlepage}
\begin{flushleft}
%       \hfill                      {\tt hep-th/1105.****}\\
       \hfill                       MPI-2011-20 \\
       \hfill                       FIT-HE-11-01 \\
       \hfill                       CCTP-2011-04  \\
\end{flushleft}
\vspace*{3mm}
\begin{center}
{\bf\LARGE Holographic (De)confinement Transitions in Cosmological Backgrounds}

\vspace*{5mm}
\vspace*{12mm}
{\large Johanna Erdmenger\footnote[1]{\tt jke@mppmu.mpg.de}, 
Kazuo Ghoroku\footnote[2]{\tt gouroku@dontaku.fit.ac.jp} and
Ren\'e Meyer\footnote[3]{\tt meyer@physics.uoc.gr}}

\vspace*{8mm}
{${}^{1}$Max-Planck-Institut f\"ur Physik (Werner-Heisenberg-Institut)}\\
F\"ohringer Ring 6, 80805 M\"unchen, Germany,\\
\vspace*{4mm}
{${}^{2}$Fukuoka Institute of Technology, Wajiro, 
Higashi-ku}\\
{Fukuoka 811-0295, Japan,\\}
\vspace*{4mm}
{${}^{3}$Physics Department, University of Crete}\\
{P.O. Box 2208, 71003 Heraklion, Crete, Greece.}

\vspace*{10mm}
\end{center}

\begin{abstract}

{For type IIB supergravity with a running axio-dilaton, we 
 construct bulk solutions which admit a cosmological background metric of Friedmann-Robertson-Walker type. 
These solutions include both a dark radiation term in the bulk as well as a four-dimensional (boundary) cosmological constant, while gravity at the boundary 
remains non-dynamical. 
We holographically calculate the stress-energy tensor, showing that it
consists of two contributions: The first one, generated by the dark
radiation term, leads to 
the thermal fluid of ${\cal N}=4$ SYM theory, while the second, the conformal anomaly, 
originates from the boundary
cosmological constant. Conservation of the boundary stress tensor
implies that the boundary cosmological constant is time-independent,
such that there is no exchange between the two stress-tensor contributions.
We then
study (de)confinement by 
evaluating the Wilson loop in these backgrounds. While the dark
radiation term favours deconfinement, a negative cosmological constant drives the system into a confined phase.
When both contributions are present, 
we find an oscillating universe with negative cosmological constant which 
undergoes periodic (de)confinement transitions as the scale of three
space 
expands and re-contracts.
}

\end{abstract}
\end{titlepage}

\tableofcontents

\newpage

\section{Introduction}\label{intro}

%%%%%%%%%%%%%
Gauge/gravity duality \cite{MGW} has proved to be extremely successful in describing 
strongly coupled systems. This applies in particular to confining theories
which can be modelled for instance using non-trivial dilaton flows
\cite{Gubser:1999pk,KS2,LT}. In these models, the
Wilson loop displays an area law. Moreover, gauge/gravity duality has also
proved useful in describing deconfined finite temperature field theories which are
naturally assumed to be dual to asymptotically AdS black holes.

Both confinement and horizon formation also arise in quantum field theories 
on curved space backgrounds, in Anti-de Sitter and de Sitter geometries,
respectively. In the gauge/gravity duality context, this has
been investigated for instance in \cite{Hawking,Alishahiha1,Alishahiha2,H,GIN1,GIN2,GSUY,AMR} by
considering a boundary cosmological constant $\lambda$ in the four-dimensional boundary
quantum field theory. 
{Holographic studies of strongly coupled quantum field theories in
  curved backgrounds are however not only interesting in their own
  right (e.g. in order to verify the properties of particle production
  phenomena such as the Unruh effect at strong coupling), but also
  from the point of view of AdS/CFT dualities for time-dependent
  backgrounds. In particular, standard cosmological evolution in the
  presence of a cosmological constant can yield de Sitter or Anti-de
  Sitter geometries. Thus from studying gauge theories in these 
  backgrounds we expect to learn about the properties of matter in the early universe (e.g. during inflation).% such as its equation of state for example.

With this situation in mind, in this paper we consider gravity duals of field theories on cosmological 
backgrounds where in the dual gravitational description a {\it bulk 
radiation term} is present in addition to a boundary cosmological constant.} 
This term has first been considered in brane world models in \cite{Kraus:1999it,Kehagias:1999vr,BDEL,Lang}. 
Due to its schematic form $C/a^4$, with $a$ being the scale factor, it corresponds to
a relativistic radiation contribution to the energy density. We discuss the interplay
between this radiation and the boundary cosmological constant in the boundary energy momentum tensor, as well as their effects on the temporal Wilson loop. {We find
that the combined effect of the dark radiation term and the boundary cosmological constant introduces an effective dynamics into the
dual field theory, triggering (de)confinement transitions for the Wilson loop. For vanishing boundary cosmological constant and flat horizon topology, the dark
radiation term gives just the Stefan-Boltzmann contribution $\rho \sim T^4$  to the 
 boundary energy density. In the other cases the relation  to the temperature is more involved due to the time-dependence of the background geometry, as we
discuss.}

As a further ingredient, we consider a running axio-dilaton similarly
to the model of Liu and Tseytlin \cite{LT}. The axio-dilaton
introduces a finite gluon condensate which on flat space leads to confinement. 

Our main result are explicit evaluations for Wilson loops in the field 
theories dual to the gravity solution with dark radiation term 
for the three cases of positive, vanishing and negative boundary cosmological 
constant. The general intuition arising from the static quark-antiquark potential is that the dark radiation term always drives the 
system into a deconfined phase (with the Wilson loop displaying a perimeter law), since it acts similarly to a temperature in flat space. We find an interesting pattern for the (de)confining behaviour of the 
Wilson loop, depending on the sign of the boundary cosmological constant:
\begin{enumerate}

\item For positive cosmological constant, the theory is always in a deconfined state, even for vanishing dark radiation term, which is in accordance with the general expectation that de Sitter-like expansion tends to destabilise bound states.

\item {For vanishing cosmological constant our bulk metrics are
    diffeomorphism equivalent to topological AdS black holes
    \cite{Tetradis1,Tetradis2}. In this case, the running axio-dilaton
    as in \cite{LT} is
    crucial for determining the confinement properties.
The Wilson loop shows confinement if the dark radiation constant is
vanishing and deconfinement otherwise, both for flat and hyperbolic
horizons. As discussed further below and in sec.~\ref{confinement},
the non-trivial dilaton flow is essential for the Wilson loop confinement in this
case, as in the absence of the running dilaton the quark-antiquark
potential would be screened for a gravity dual with 
hyperbolic topological black hole \cite{LL} at all temperatures.
}

\item For negative cosmological constant, we find an interesting (de)confinement transition which occurs due to the competition between the deconfining dark radiation term and the confining nature of Anti-de Sitter-like contraction: For small scale factors the Wilson loop is deconfined, while for large scale factors it is confined. Intuitively this can be thought of as the Wilson loop probing the (holographically defined) field theory vacuum whose energy-momentum VEV now has two components - the conformal anomaly component due to the boundary cosmological constant, and the dark radiation component which in essence behaves like thermal relativistic radiation, getting diluted by the scale factor as $a^{-4}(t)$.\footnote{This interpretation of the dark radiation constant has also been given in the gauge/gravity context in \cite{Kiritsis:2005bm,Tetradis1,Tetradis2}. Furthermore, the works \cite{Tetradis1,Tetradis2} write the AdS-Schwarzschild black hole (without a running dilaton) in a cosmological foliation, which is possible in our construction as well. However, our setup is more involved due to the presence of the boundary cosmological constant and the running dilaton (see sec.~\ref{confinement}). Finally, a related study \cite{Brito} for $k=0$ and without dark radiation found an interplay between boundary cosmological constant and the tension of an IR brane sourcing the geometry.} Hence close to the singularity the dark radiation component dominates, driving the system to deconfinement, while away from it the confining nature of the negative cosmological constant dominates.

\end{enumerate}
{
In this work we mainly focus, for simplicity, on the Wilson loop as a
measure of quark-antiquark (de)confinement. On curved space-times,
other measures of confinement such as the density of states or the
mass gap criterion do not necessarily coincide with the Wilson loop
criterion: For example, in \cite{CW,HMR} it was argued that the Wilson
loop confines on AdS spaces at any temperature, while \cite{AMR}
showed that Neumann boundary conditions on the boundary of AdS space
allow for a large N deconfinement transition at finite temperature,
with the deconfined phase being characterised by a $O(N^2)$ density of
states at low energies. We hence leave a thorough investigation of the
subtleties involved in relating these different criteria for
confinement for future work, and rather focus on the Wilson loop as a criterion
to characterize our holographic backgrounds. The reader should also
consult sec.~\ref{confinement} for a more in-depth discussion of the
case of vanishing cosmological constant and hyperbolic horizon: In this case the Wilson loop and the density of states measure indeed do not agree, due to the presence of the gluon condensate (i.e. the running dilaton) and since the thermodynamical contributions of the axio-dilaton cancel each other in the Liu-Tseytlin ansatz.
}

On the technical side, we  decouple the axio-dilaton dynamics from the
five-dimensio\-nal metric in the same way as in \cite{LT}. 
We then solve the Einstein equations of five-dimensional Einstein-Hilbert gravity 
with a cosmological Ansatz for the metric already used in \cite{BDEL} in the context of brane-world cosmology. The Friedmann equation arises from the constraint equation of the bulk Einstein equations, and we include the dark radiation term into our analysis. 
By imposing the usual Dirichlet boundary conditions of holography, the four-dimensional boundary gravity remains 
non-dynamical, as there is no four-dimensional 
Einstein-Hilbert action on the boundary. Besides the dark radiation term we also allow for a boundary cosmological constant in our Friedmann equation. We find that requiring boundary diffeomorphism invariance 
leads to a time-independent boundary cosmological constant. 

The remainder of this paper is structured as follows: In the next section, the
holographic background is given and we discuss how the 
boundary cosmological constant arises. In section 3, the holographic
interpretation of the dark radiation term in our approach is
illuminated using the specific example of vanishing boundary
cosmological constant. Section 4 discusses the solution for
finite boundary cosmological constant, and how boundary diffeomorphism
invariance (i.e. conservation  of the boundary stress-energy tensor) forces the cosmological constant to be actually time-independent. 
Section 5 then derives the main result of this paper: The Wilson loop expectation values are calculated and their (de)confinement properties are classified. Summary and discussions are given in the final section 6.

\section{The Background Geometry}\label{background}

{In this section we first review the reduction of ten-dimensional type
  IIB supergravity to a five-dimensional dilaton gravity by a
  Freund-Rubin Ansatz which allows for a nontrivial axio-dilaton. This
  ansatz, first employed in \cite{KS2,LT,GGP}, links the axion with
  the dilaton in a way which allows to describe $1/4$ supersymmetric
  D3-D(-1) solutions. Supersymmetry can then be broken by introducing
  finite temperature. In \ref{sec22} we then solve the five-dimensional Einstein equations with a time-dependent ansatz for the metric along the lines of \cite{BDEL} and find holographic backgrounds describing a cosmological evolution at the boundary. In this course we identify the boundary cosmological constant, driving the cosmological evolution of the boundary metric, when solving the constraint equations in the bulk.
}

\subsection{Five-dimensional Dilaton Gravity from IIB Supergravity}\label{sec21}

We start from the 
ten-dimensional type IIB supergravity retaining the dilaton
$\Phi$, axion $\chi$ and selfdual five form field strength $F_{(5)}$,
\beq
 S={1\over 2\kappa^2}\int d^{10}x\sqrt{-g}\left(R-
{1\over 2}(\partial \Phi)^2+{1\over 2}e^{2\Phi}(\partial \chi)^2
-{1\over 4\cdot 5!}F_{(5)}^2
\right), \label{10d-action}
\eeq
where other fields are consistently set to zero,  and 
$\chi$ is Wick rotated \cite{GGP}.
Under the Freund-Rubin
ansatz for $F_{(5)}$, 
$F_{\mu_1\cdots\mu_5}=-\sqrt{\Lambda}/2~\epsilon_{\mu_1\cdots\mu_5}$ 
\cite{KS2,LT}, and for the 10d metric taken as $M_5\times S^5$, 
$$ds^2_{10}=g_{MN}dx^Mdx^N+g_{ab}dx^adx^b\, ,\quad M,N=0,\dots,4\,,\quad a,b=5,\dots,9\,,$$ 
the equations of motion of the non-compact five-dimensional part
$M_5$ become 
\footnote{The five-dimensional part $M_5$  of the
solution is obtained by solving the following reduced 
Einstein frame 5d action,
\beq
 S={1\over 2\kappa_5^2}\int d^5x\sqrt{-g}\left(R+3\Lambda-
{1\over 2}(\partial \Phi)^2+{1\over 2}e^{2\Phi}(\partial \chi)^2
\right)\,. \label{5d-action}
\eeq
The opposite sign
of the kinetic term of $\chi$ is due to the fact that
the Euclidean version is considered here \cite{GGP}.}
\beq\label{metric}
 R_{MN}={1\over 2}\left(\partial_M\Phi\partial_N\Phi-
         e^{2\Phi}\partial_M\chi\partial_N\chi\right)-\Lambda g_{MN}
\eeq
\beq \label{p}
 {1\over \sqrt{-g}}\partial_M\left(\sqrt{-g}g^{MN}\partial_N\Phi\right)=
    -e^{2\Phi}g^{MN}\partial_M\chi\partial_N\chi\ , 
\eeq 
\beq\label{chieq}
   \partial_M\left(\sqrt{-g}e^{2\Phi}g^{MN}\partial_N\chi\right)=0
\eeq
These equations have a supersymmetric solution when the following ansatz is
imposed for the axion $\chi$ \cite{GGP,KS2},
\beq\label{ansa1}
     \chi=-e^{-\Phi}+\chi_0 \ .
\eeq
In this case, 
using the ansatz (\ref{ansa1}) in (\ref{metric})--(\ref{chieq}) gives rise to the two equations
\beq\label{gravity}
 R_{MN}=-\Lambda g_{MN}
\eeq
and
\beq\label{phi}
   \partial_M\left(\sqrt{-g}g^{MN}\partial_N e^{\Phi}\right)=0 \,, 
\eeq
where (\ref{p}) and (\ref{chieq}) now may be shown to coincide using
(\ref{phi}). The latter set of equations is also useful for finding
finite temperature solutions in which supersymmetry is broken.

\subsection{Solution with Dark Radiation}\label{sec22}

We examine here time-dependent solutions which include a ``dark radiation" term
\cite{BDEL,Lang} (also known as `mirage energy density' \cite{Kiritsis:2005bm}).
To find this term, we change the
radial coordinate $r$ to $y$, where $r/R=\mu r=e^{\mu y}$ and 
$\mu=1/R=\sqrt{\Lambda}/2$, and 
we consider the following Einstein frame metric,
\beq
ds^2_{\rm E}=-n^2(t,y)dt^2+a(t,y)^2\gamma_{i,j}dx^idx^j+dy^2\,,\quad i,j=1,\dots,3. 
\label{y-co}
\eeq
In this metric, we obtain from the Einstein equation for the
$tt$ and $yy$ components \cite{BDEL} 
\beq\label{cosmo}
 \left({\dot{a}\over na}\right)^2+{k\over a^2}=
   -{\Lambda\over 4}+\left({a'\over a}\right)^2
  +{C\over a^4}\ ,
\eeq
where $\dot{a}=\partial a/\partial t$ and $a'=\partial a/\partial y$. Note that this is a first order equation, 
integrated from the second order Einstein equations (see \cite{BDEL} for their explicit form). 
It then turns out that without any additional matter in the bulk, the integration constant $C$ must be a constant with respect to both $y$ and
$t$ in order to satisfy both the $tt$ and $yy$ components of Einsteins
equations. This constant $C$ 
appears in the equation (\ref{cosmo}) in the form
${C\over a^4}$, which is usually referred to as ``dark radiation" term, since it behaves exactly as a 
component of relativistic radiation which, in the context of braneworld models, 
leaks from the bulk into the UV brane \cite{Lang}.

It also needs to be checked whether the Bianchi identities and the
  $ij$ and  $ty$ components of Einsteins equations are satisfied with the above Ansatz. As shown in \cite{BDEL}, the first two are satisfied upon the use of eq.~\eqref{cosmo}, while the latter relates the free function $n(t,y)$ to $a(t,y)$ up to a time-dependent integration constant,
\beq\label{tyequation}
0 = \frac{n'}{n}\frac{\dot a}{a} - \frac{\dot a'}{a}\,.
\eeq
This last equation is solved 
by setting the following ansatz \cite{BDEL,Lang},
\beq\label{eqn}
   n(t,y)={\dot{a}(t,y)\over \dot{a}_0(t)}\, , \quad a=a_0(t) A(t,y)\, .
\eeq
Then the equation for $A(t,y)$ is obtained from (\ref{cosmo}) as
\beq\label{A}
 \left({\dot{a}_0\over a_0}\right)^2+{k\over a_0^2}=
   -{\Lambda\over 4}A^2+\left({A'}\right)^2
  +{C\over a_0^4 A^2}\ , 
\eeq
where $A'=\partial A/\partial y$. {Looking at eq.~\eqref{A}, we recognize its left-hand side as part of the Friedmann equation from standard cosmology. More precisely, it is the part of Friedmann's equation without the cosmological constant term. In particular, since the left-hand side of eq.~\eqref{A} is a function of time only, the right-hand side of \eqref{A} must also be only a function of time, i.e. independent of the radial coordinate $y$. The right-hand side of \eqref{A} thus effectively acts as a time-dependent vacuum energy ``source term'' for the cosmological evolution at the boundary, described by the left-hand side. Introducing a time-dependent boundary cosmological ``constant'' $\lambda(t)$, we can thus separate eq.~\eqref{A} into its left and right hand sides, yielding two independent equations. This procedure is similar to separation of variables when solving differential equations: For general time-dependent $\lambda(t)$ the above replacement of eq.~\eqref{A} by the two equations \eqref{Friedman} and \eqref{A2nd} does not affect the solution space, since every solution of \eqref{Friedman} and \eqref{A2nd} will be a solution of \eqref{A}, and \textit{vice versa}. 

Doing so, the left hand side of (\ref{A}) then becomes the four-dimensional
Friedmann equation with
a four-dimensional boundary cosmological 
term   $\lambda(t)$,}
\beq\label{Friedman}
   \left({\dot{a}_0\over a_0}\right)^2+{k\over a_0^2}=\lambda(t)\, , 
\eeq
where $k=\pm 1$ or 0. From standard cosmology, \eqref{Friedman} only yields universes with spherical ($k=+1$) topology for $\lambda>0$, while for $k = 0$ the allowed choices are $\text{sgn} \lambda=0,+1$, and for negative spatial curvature $k=-1$ even a spatially homogeneous and isotropic universe of constant negative curvature is allowed, i.e. $\text{sgn} \lambda=-1,0,+1$ are possible choices. 

For any $\lambda(t)$, 
$A(t,y)$ can then be solved for by the following first-order differential equation in the variable $y$,
\beq\label{A2nd}
 \lambda(t)=
   -{\Lambda\over 4}A^2+\left({A'}\right)^2
  +{C\over a_0^4 A^2}\ , 
\eeq
using the solution $a_0(t)$ of (\ref{Friedman}). {In the above treatment of eq.~\eqref{A} we introduced an \textit{a priori} time-dependent function $\lambda(t)$. In an evolving universe, a time-dependent cosmological constant however would have to be sourced by additional energy-momentum sources at the boundary or in the bulk, which generate the relevant piece in the energy-momentum tensor that ensures energy-momentum conservation. Since in the holographic context with standard Dirichlet boundary condition, gravity at the boundary is not dynamical (i.e. the background metric for the dual field theory is a fixed background field), no boundary matter source can influence the boundary  metric. The holographic energy-momentum tensor itself thus has to be conserved. We will calculate the holographic energy-momentum tensor in sec.~\ref{sec4}, but quote the result here already and argue that stress-energy conservation forces the cosmological constant to actually be time-independent.

The general solution of eq.~\eqref{A2nd}, which will be analysed in more detail in sec.~\ref{sec4}, is
\beq
A = \frac{r}{R} \left( \left(1 - \frac{\lambda(t)R^2}{4} \frac{R^2}{r^2}\right)^2 + \frac{C R^2}{4 a_0^4(t)} \frac{R^4}{r^4} \right)^{\frac{1}{2}}\,.
\eeq
Using standard holographic techniques (for details see sec.~\ref{442}), the vacuum expectation value of the boundary stress-energy tensor is found to be of perfect fluid form,
\beqa
\langle T^\mu{}_\nu \rangle &=& \text{diag}(-\rho,p,p,p)\,,\quad  \alpha = \frac{4R^3}{16\pi G_N^{(5)}}\,,\\
\rho &=& 3 \alpha \left( \frac{C }{4R^2 a_0^4(t)} + \frac{\lambda(t)^2}{16} \right)\,,\quad p = \alpha \left( \frac{C }{4R^2 a_0^4(t)} - \frac{3\lambda(t)^2}{16} \right)\,.
\eeqa
If we now impose the holographic stress-energy tensor to be conserved, $\nabla_\mu \langle T^\mu{}_\nu \rangle = 0$, we actually require a continuity equation for pressure and energy density,
\beq
0 = \dot \rho + 3 H(\rho + p)\,.
\eeq
In order to satisfy this continuity equation, the boundary cosmological constant $\lambda(t)$ then has to be a constant,
\beq
\dot \lambda(t)=0\,.
\eeq
Physically speaking, requiring the stress-energy tensor to be conserved amounts to requiring the holographically defined generating functional to be invariant under boundary diffeomorphisms. The conservation equation $\nabla_\mu \langle T^\mu{}_\nu \rangle = 0$ then is the one-point function diffeomorphism Ward-Takahashi identity. In holographic renormalisation this is a natural outcome since both the regularised on-shell action as well as the counterterms are constructed in a manifestly boundary diffeomorphism invariant way. {A possible non-conservation of the bulk part of $\langle T_{\mu\nu} \rangle$ can then only be cancelled by additional boundary terms which change the chosen boundary conditions from Dirichlet to Neumann or mixed ones \cite{Marolf}, hence inducing additional dynamical degrees of freedom into the boundary theory.} We thus conclude that to constitute a physically meaningful holographic background, the boundary cosmological constant must be an actual constant in time. One should note that restricting the solution space of eqs.~\eqref{A} or \eqref{Friedman},\eqref{A2nd} does not influence the argument given above concerning the equivalence of the (restricted) solution space of both sets of equations.
}

{We will show in the following (in particular in section~\ref{confinement}) that 
the solution $A(t,y)$ of this equation 
encodes important dynamical properties of the gauge theory in a Friedmann-Robertson-Walker
universe. 
As will be shown in section~\ref{confinement}, the behaviour of Wilson loop expectation values as calculated from a minimal string world sheet, and hence the (de)confinement properties of the vacuum show an interesting competition between the dark radiation constant 
 $C > 0$, and the boundary cosmological constant $\lambda$.}

{Finally, we would like to comment on the physical situation concerning the dark radiation term and the boundary cosmological constant  in brane world models \cite{RS1,bre}, which is slightly different. In these models, due to the fact that the UV brane sits at a finite cutoff and due to the chosen boundary conditions, an energy exchange between the bulk and the brane is possible.  In particular, the value of $\lambda$ is always tuned by the bulk metric and the five-dimensional cosmological constant $\Lambda$, and the dark radiation term ${C\over a_0^4 A^2}$ is considered to be an energy flux between 
 brane and bulk. In the holographic setup we consider here, due to the standard Dirichlet boundary conditions chosen in holography, no bulk-boundary energy exchange is possible, and the boundary cosmological constant can be freely tuned. The holographic setup is thus less rigid compared to the brane-world models.
}

\section{Holographic Interpretation of Dark Radiation}\label{ZeroLambda}

Above we have shown how to obtain a consistent background
  solution describing a boundary metric undergoing cosmological
  evolution under the influence of both a boundary cosmological
  constant and a dark radiation term in the bulk. Here we
  concentrate on the holographic interpretation of the dark radiation
  term for the simplest case, i.e. the case of vanishing boundary
  cosmological constant. We find that the dark
  radiation term introduces a temperature for the boundary ${\cal
    N}=4$ field theory.

\subsection{Solution for Vanishing Boundary Cosmological Constant}

For the case of vanishing boundary cosmological constant, a solution
of (\ref{Friedman}) is given by 
$\lambda=0$, $k=0$ and $a_0(t)=1$, and $A(t,y)=A(y)$ is
obtained by solving (\ref{A2nd}) with $\lambda=0$.
This gives
\beq
  A=e^{\mu y}\left(1+\tilde{c}_0 e^{-4\mu y}\right)^{1/2} \, ,
\eeq
where $\tilde{c}_0=C/(4\mu^2a_0^4)=C/(4\mu^2)$ since $a_0=1$. 
From Eq. (\ref{eqn}), we obtain
\bea
 n&=&A-{1\over A}{2C\over \lambda }
       e^{-2\mu \tilde{y}} \nonumber \\ 
   &=&e^{\mu y}{1-\tilde{c}_0 e^{-4\mu y}\over \sqrt{{1+\tilde{c}_0
         e^{-4\mu y}}}} \, .
\eea
Then, using $r/R=e^{\mu y}$
the full Einstein metric is given by  
\beq
ds^2_{10}={r^2 \over R^2}\left(-\bar{n}^2dt^2+\bar{A}^2(dx^i)^2\right)+
\frac{R^2}{r^2} dr^2 +R^2d\Omega_5^2 \,,
\label{finite-c-sol-2}
\eeq
\beq
 \bar{A}=\left(1+\tilde{c}_0 \left(R\over r\right)^{4}\right)^{1/2}\, , \quad
\bar{n}={1-\tilde{c}_0 \left(R\over r\right)^{4}\over 
       \sqrt{{1+\tilde{c}_0 \left(R\over r\right)^{4}}}}\,.
\eeq
For the dilaton (\ref{phi}) we find,  using the above metric,
\beq\label{dilaton-C}
 e^{\Phi}=1+{q\over 2\tilde{c}_0R^4}\log{1+\tilde{c}_0 (R/r)^4
\over 1-\tilde{c}_0 (R/r)^4}\, .
\eeq
Here the integral constant $q$ corresponds to the gauge condensate
$\langle \tr F^2 \rangle$,
 and the 
boundary condition $e^{\Phi}\to 1$ for $r\to\infty$ is imposed. 
For $\tilde{c}_0=0$, this solution reduces to the supersymmetric one used in 
\cite{GY}, and $e^{\Phi}$ diverges at $r=0$ if $\tilde c_0 >0$.

\subsection{Holographic Interpretation of Dark Radiation}

In order to give an interpretation to 
dark radiation constant $C$, we rewrite the solution eq.~\eqref{finite-c-sol-2} as a planar AdS-Schwarzschild black hole.  
The five-dimensional part of the metric in the Einstein frame 
can be brought into that form, 
\bea\label{thermal}
 ds^2_{(5)}&=&{{\tilde r}^2 \over R^2}\left(-f({\tilde r})dt^2 + (dx^i)^2\right)+
\frac{R^2d{\tilde r}^2}{{\tilde{r}}^2 f({\tilde r})}\,,\quad f(\tilde{r}) = 1 - \frac{{\tilde r}_0^4}{{\tilde r}^4}\,, 
\eea
by the coordinate redefinition
de\beq\label{thermal-1}
{\tilde r} = r \sqrt{1 + \frac{R^4}{r^4} {\tilde c}_0}\quad\Rightarrow {\tilde r}_0 = (C R^6)^{1/4}\,.
\eeq
Thus the dark radiation constant $C$ has to be positive, and sets the horizon radius of the AdS-Schwarzschild black hole. The dark radiation constant is nothing but the mass of the AdS-Schwarzschild black hole and, applying standard holographic renormalisation \cite{KSS,BFS,FG} we find the dual stress-energy tensor to be of perfect fluid form
\beq
 \langle T_{\mu\nu}^{(0)}\rangle={{C R} \over 16\pi G_N^{(5)}}
\text{diag}(3,1,1,1)\,.
\eeq\label{stress0}
We thus conclude that in holography the dark radiation constant defines a temperature for the fields of the dual field theory. For a non-expanding cosmology ($k=\lambda=0$ as in this case) this directly leads to a field theory (in this case ${\cal N}=4$ with gluon and instanton condensate) at finite (Hawking) temperature
\beq\label{temperature}
 T_{H_0}={(4\tilde{c}_0)^{1/4}\over \pi R}\,.
\eeq
Using this temperature and the energy momentum tensor (\ref{stress0}), we can in particular confirm the 
 Stefan-Boltzmann law for the energy density\footnote{The background metric in this special case is just the flat Minkowski metric $\eta_{\mu\nu} = \text{diag}(-1,1,1,1)$.} $\rho = - \langle T^0{}_0 \rangle$,
\beq\label{StefanBoltzmann}
 \rho={4R^3\over 16\pi G_N^{(5)}}\left(3{\tilde{c}_0\over R^4}\right)
  ={3N^2\over 8}\pi^2T_H^4\,.
\eeq
where we used $G_N^{(5)}=8\pi^3{\alpha'}^4g_s^2/R^5$ and $R^4=4\pi N{\alpha'}^2g_s$. 
This expression reproduces the known results of \cite{GKP,NS}. 
In brane-world models, the dark radiation term has been interpreted as the radiation of the bulk gravitons 
which transfer the energy of the fields in the brane to the bulk. In
\cite{Gubser,Lang} it was noted that the dark radiation constant
corresponds to the mass of the bulk AdS-Schwarzschild black hole. In
their contexts, gravity is dynamical on the UV brane, and the dark
radiation term appears
in the Friedmann equation on the brane. As noted already in
section~\ref{background}, the holographic setup considered in this
work is different due to the Dirichlet boundary conditions imposed. 
Here, gravity is not dynamical at the boundary of space-time. Instead, 
in our case the dark radiation term is dual to
the energy density of the ${\cal N}=4$ $U(N)$ SYM fields in a thermal state, as evident from the Stefan-Boltzmann law \eqref{StefanBoltzmann}. The dark radiation constant $C$, which appears as an integral constant when solving Einsteins equations \cite{BDEL}, sets the temperature of the dual field theory. To the best of our knowledge such a holographic interpretation of the bulk radiation term has not yet been given in the literature before. This interpretation will qualitatively also hold in the time-dependent cosmologies considered in sections~\ref{sec4}~and~\ref{confinement}: We find that in all cases the dark radiation constant contributes in a thermal manner to the holographic stress-energy tensor of the system, with a time-dependent prefactor $a_0(t)^{-4}$ associated with the dilution of relativistic radiation due to expansion or contraction of the (boundary) universe. On the other hand, the boundary cosmological constant yields a conformal anomaly contribution to the stress-energy tensor. We will see that both contributions can compete, giving rise to interesting dynamics.

\section{Holography for Boundary $\mathbf{(A)dS_4}$ Space-Times}\label{sec4}\label{sec:4}

Above we saw that for vanishing boundary cosmological
constant, the dark radiation constant corresponds to a temperature for
the ${\cal N} = 4$ fields. In this section we treat the case of finite
boundary cosmological constant, and discuss in particular the boundary
stress-energy tensor. 
We show that the dark radiation term induces a relativistic
radiation contribution to the boundary stress-energy tensor, varying in time
with the well-known $a_0(t)^{-4}$ dependence during cosmological expansion.
Furthermore, we find that stress-energy conservation in the boundary theory
forces the boundary cosmological constant to be time-independent.

\subsection{Solution for Finite Boundary Cosmological Constant}

A solution of \eqref{A2nd} for finite $\lambda(t)$ is 
\beq
  A=e^{\mu y}\left(\left[1-{\lambda(t)\over 4\mu^2}e^{-2\mu y}\right]^2
+\tilde{c}_0(t) e^{-4\mu y}\right)^{1/2} \, ,
\eeq
where $\tilde{c}_0=C/(4\mu^2a_0(t)^4)$. Here we have chosen asymptotic boundary conditions 
\beq
 A(y=\infty )=e^{\mu y}=r/R \, ,
\eeq
where $\mu=1/R$, i.e. we require the asymptotic form of the metric 
to be AdS$_5$. We find that  $A$ has time-dependence through $a_0(t)$ in $\tilde{c}_0$ and also $\lambda(t)$. This point is important to determine the structure of the metric below.

From Eq. (\ref{eqn}), we obtain
\beq
 n={e^{2\mu y}\over A(t,y)}\left(\left[1-{\lambda(t)\over 4\mu^2}e^{-2\mu y}\right]^2
-\tilde{c}_0(t) e^{-4\mu y}\right)\,.
\eeq
With $r/R=e^{\mu y}$, the full Einstein frame metric is then given by  
\beq
ds^2_{10}={r^2 \over R^2}\left(-\bar{n}^2dt^2+\bar{A}^2a_0^2(t)\gamma^2(x)(dx^i)^2\right)+
\frac{R^2}{r^2} dr^2 +R^2d\Omega_5^2 \, ,
\label{finite-c-sol-3}
\eeq
where
\beq\label{abarnbar}
 \bar{A}=\left(\left(1-{\lambda\over 4\mu^2}\left({R\over r}\right)^2\right)^2+\tilde{c}_0 \left({R\over r}\right)^{4}\right)^{1/2}\, , \quad
\bar{n}={\left(1-{\lambda\over 4\mu^2}\left({R\over r}\right)^2\right)^2-\tilde{c}_0 \left({R\over r}\right)^{4}\over 
       \sqrt{\left(1-{\lambda\over 4\mu^2}\left({R\over r}\right)^2\right)^2+\tilde{c}_0 \left({R\over r}\right)^{4}}}\,.
\eeq
{The above metric has no naked singularities for time-independent $\lambda$, as we checked by calculating $R$, $R_{\mu\nu} R^{\mu\nu}$ and the Kretschmann scalar $R_{\mu\nu\rho\sigma} R^{\mu\nu\rho\sigma}$.}\footnote{{Recently it was noted in a similar but not identical construction \cite{LiPang} that naked singularities might appear when deviating from pure dS expansion. Their singularity so far cannot be shielded by a horizon. In contrast, the Einstein frame curvature singularities in the backgrounds considered here are always behind the horizon $g_{tt}=0$, and coincide with the cosmological singularities $a_0(t)=0$.}} We will see in sec.~\ref{4.4} that $\lambda$ also needs to be time-independent in order to ensure boundary energy-momentum conservation. 
Note that we use a coordinate system in which the constant curvature three-space has the metric
\beq\label{3space}
d \Omega_k^2 = \frac{\extd \vec{x}^2}{(1+k \vec{x}^2/4)^2}\,.
\eeq
These are simply the standard spherical coordinates on the isotropic and homogenous three-space, with a conformal factor.

\subsection{Almost Constant Scale Factor  and Adiabatic Expansion}

Quantum fields in an expanding space usually are not in thermal equilibrium, not even locally, unless the expansion rate is slow compared to the equilibration time of the system. This should be the case for very small but nonzero boundary cosmological constant $\lambda$, in which case the scale factor $a_0(t)$ would still be changing with time, but with a very slow rate. In other words, the Hubble rate $H = \dot a_0 / a_0$ is small. In this case we can still make statements about the ``slowly varying'' temperature of the system, corresponding to the adiabatic regime.\footnote{The system evolves adiabatically, starting from $t_0$, roughly for a time span $\sqrt{|\lambda|} |t-t_0| \ll 1$.} {We find the horizon as the zero of the $g_{tt}$ metric coefficient in \eqref{finite-c-sol-3}, which is at 
\beq\label{horizon}
   r_H=R\sqrt{\tilde{c}_0^{1/2}+{{\lambda}\over 4\mu^2}}
\eeq
for $\lambda > -(4\mu^2){\tilde{c}_0}^{1/2} = \lambda_c$. 
If the scale factor $a_0$ is slowly changing, it is possible to approximately  satisfy \eqref{Friedman} with a time-independent $a_0$, by taking $k=1$ for $\lambda >0$ and $k=-1$ for $\lambda < 0$ in the Friedmann equation (\ref{Friedman}), 
\beq
  a_0\approx 1/|\lambda|^{1/2}\, , \quad
  \gamma(x)= \left(1+ k {x_i^2\over 4}\right)^{-1}\,.
\eeq
In this case, $\partial_{\tau}r_H \approx 0$, and from the near-horizon geometry 
\beq
 ds^2\simeq 8\left({r_H\over R}\right)^2\epsilon^2 d\tau^2
     +R^2d\epsilon^2+\cdots\, .
\eeq
a (slowly varying) Hawking temperature can be found for $\lambda > -(4\mu^2)\sqrt{\tilde{c}_0}$, reading 
\beq
 T_H={\sqrt{\lambda/(2\mu^2)+(4\tilde{c}_0)^{1/2}}\over \pi R}\,.
 \label{temperature2}
\eeq
We thus find that negative (positive) $\lambda$ decreases (increases) the effective temperature for the dual field theory. Furthermore we observe that the regime $\lambda<-(4\mu^2)\sqrt{\tilde{c}_0}$ is special: Formally, the Hawking temperature calculation does not apply to that case even if dark radiation is present, since $g_{tt}$ has no real zero any more, i.e. there is no horizon. We will see in sec.~\ref{confinement} that in this regime the Wilson loop shows a confining area law behaviour. The situation is thus similar to  the Sakai-Sugimoto model \cite{SS}, where the gravity dual of the confined phase is a cigar-shaped geometry which smoothly caps off instead of admitting a black hole horizon. 
}

\subsection{Dilaton Solution}

{In addition to the metric considered above, the other important
  field in our system is the running dilaton, whose exponential is related to the gauge coupling in the dual field theory.} The solution to the dilaton equation of motion reads
\bea\label{dilaton-lam}
 e^{\Phi}&=&{q\over 2\tilde{c}_0\left(1+{\lambda^2\over 16\tilde{c}_0\mu^4}\right)}
\left\{
\log{1+\tilde{c}_0(R/r)^4+(\lambda R/(4\mu^2r))^2((R/r)^2-8\mu^2/\lambda)
  \over 1-\tilde{c}_0 (R/r)^4+(\lambda R/(4\mu^2r))^2((R/r)^2-8\mu^2/\lambda)}
\right.\nonumber \\
  &+&\left.
  {\lambda\over 2\tilde{c}_0^{1/2}\mu^2}(\tan^{-1}\beta+\tanh^{-1}\beta-{1-i\over 2}\pi)
\right\}+\gamma\, ,
\eea
where $q$ and $\gamma$ are the integration constants and
\beq
 \beta={(r/R)^2-\lambda/(4\mu^2)\over \tilde{c}_0^{1/2}}
\eeq

\begin{figure}[htbp]%[H]
\begin{center}
\includegraphics[width=7.0cm,height=6cm]{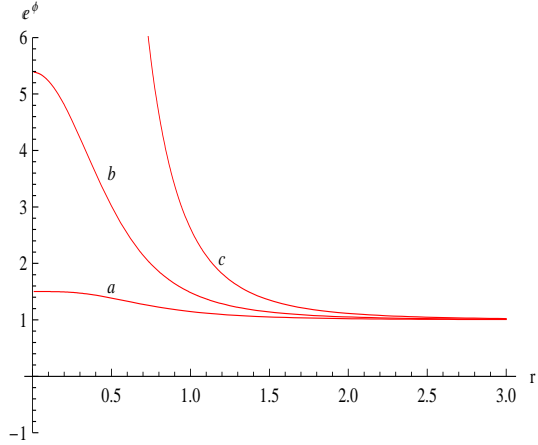}
\caption{Plots of $e^{\phi}$ vs $r$ for (a) $\lambda=-1-4\mu^2\tilde{c}_0^{1/2}$,
(b) $\lambda=-4\mu^2\tilde{c}_0^{1/2}$ and
(c) $\lambda=1-4\mu^2\tilde{c}_0^{1/2}$. Cases (a) and (c) are taken
as examples for $\lambda < -4\mu^2\tilde{c}_0^{1/2}$ and $\lambda >
-4\mu^2\tilde{c}_0^{1/2}$, respectively. Other parameters are set as $1/\mu=R=1$,
$q=2$ and $\tilde{c}_0=0.1$. In the case of (c), $e^\phi$ diverges at the horizon
$r_H=R\sqrt{\tilde{c}_0^{1/2}+{\lambda\over 4\mu^2}}$, which is 0.5 in this case. 
Note that in general $\tilde{c}_0$ is explicitly time-dependent, these curves 
represent snapshots of the dilaton solution at constant time.
\label{WL-t}}
\end{center}
\end{figure}

\begin{figure}[htbp]
\begin{center}
\includegraphics[width=10.0cm,height=5cm]{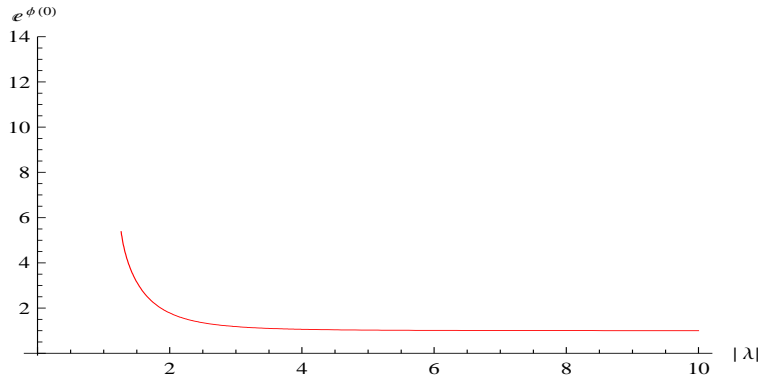}
\caption{Plot of $e^{\phi (0)}$ vs $|\lambda|$,
where the parameters are set as in Fig.~\ref{WL-t}, yielding $4\mu^2\tilde{c}_0^{1/2}=1.265$. We notice that
$e^{\phi (0)}$ is real for $|\lambda| > 4\mu^2\tilde{c}_0^{1/2}=1.265$, which corresponds to case (a) in Fig.~\ref{WL-t}, with no horizon present. For $|\lambda|<4\mu^2\tilde{c}_0^{1/2}=1.265$, case (c) of Fig.~\ref{WL-t}, the dilaton becomes complex in the region hidden behind the horizon. 
Moreover, $e^{\phi (0)}\to 1$ for $|\lambda|\to \infty$. 
{Note that since in general $\tilde{c}_0$ is explicitly
  time-dependent, this curve  
should be understood at a given instance in time.}
\label{phi0}}
\end{center}
\end{figure}

We notice the following points for the above solution (\ref{dilaton-lam}):
\begin{enumerate}

\item The above expression (\ref{dilaton-lam}) seems to be complex due to the factor
${1-i\over 2}\pi$ in the second line of (\ref{dilaton-lam}).
However, this is necessary to cancel the
imaginary part of $\tanh^{-1}\beta$ which has a constant
imaginary part $i\pi/2$ for $\beta>1$. 
The condition of $\beta>1$ is realized for $r>r_H$ (all $r>0$) in the case of 
$\lambda >-4\mu^2\tilde{c}_0^{1/2}$ ($\lambda <-4\mu^2\tilde{c}_0^{1/2}$), and hence we only give the 
 solution of $e^{\Phi}$ in this regime, and explicitly display the factor 
${1-i\over 2}\pi$. 
As a result, the above expression (\ref{dilaton-lam})
is real. 

\item The factor arctanh$(\beta)$ diverges for $\beta\to 1$, which
is realized for $r\to r_H=R\sqrt{\tilde{c}_0^{1/2}+{\lambda\over 4\mu^2}}$. The same logarithmic
divergence comes from the first logarithmic term in the equation (\ref{dilaton-lam}). 
This divergence can be seen in case (c) of Fig. \ref{WL-t}.

\item For the case of $\lambda \leq -4\mu^2\tilde{c}_0^{1/2}$, the solutions 
extend to $r=0$ and there is no divergence at any point in radial direction. The value at $r=0$
is given by 
\bea\nonumber
e^{\Phi(0)}&=&\gamma+{8q\mu^4\over \lambda^2+16\mu^4\tilde{c}_0}
\left( \log{\lambda^2+16\mu^4\tilde{c}_0\over \lambda^2-16\mu^4\tilde{c}_0} \right.\\
&&\hspace{3cm} +\left.{\lambda\over 2\tilde{c}_0^{1/2}\mu^2}(\tan^{-1}\beta_0+\tanh^{-1}\beta_0-{1-i\over 2}\pi )\right )\, ,
\label{DilatonIR}
\eea
where $\beta_0=\lambda/(4\tilde{c}_0^{1/2}\mu^2)$. 
The important point is that 
$e^{\Phi(0)}$ is finite. 
We plot its numerical value in Fig.~\ref{phi0} 
for an appropriate parameter set as a function of $|\lambda|$. 

It is interesting 
to note that the asymptotic value of $e^{\Phi(0)}$ at $\lambda=\pm\infty$ is given by $\gamma$. 
{Thus, in the limit of asymptotically large positive or negative cosmological constants, there is no running of the coupling due to the gluon condensate, but only due to the conformal anomaly induced by the background. 
 It would be interesting to further investigate this fact from a field-theoretic point of view.
}

\item Further, if we set $\gamma=1$, we obtain the following asymptotic form
\beq
e^{\Phi}\simeq 1+q/r^4+\cdots
\eeq
as ${r\to \infty}$. {This is the standard AdS/CFT expansion for a scalar dual to a $\Delta=4$ operator, which in this case is the gluon condensate $\Tr F^2$. The integration constant $\gamma$ corresponds to the nonnormalisable mode, while $q$ encodes the vacuum expectation value $\langle \Tr F^2\rangle$.}
\end{enumerate}

{
Thus, while the ultraviolet behaviour of $e^{\Phi}$ does not depend on $\lambda$, the behaviour near the infrared region is very sensitive to
the boundary cosmological constant.
For $\lambda >-(4\mu^2)\sqrt{\tilde{c}_0}$, $e^{\Phi}$ diverges at the horizon of the 
black hole configuration. On the other hand, 
for $\lambda \leq -(4\mu^2)\sqrt{\tilde{c}_0}$, $e^{\Phi}$ 
approaches a constant at
$r=0$, and $\partial_r e^{\Phi}|_{r=0}=0$.
In this case, then, the Yang-Mills coupling constant reaches at an IR
fixed point for $\lambda \leq -(4\mu^2)\sqrt{\tilde{c}_0}$. We should however note that conformal invariance of the boundary theory
is still broken due to the gravity contribution to the conformal anomaly. 
However, this will not affect on the renormalization group equation for the
Yang-Mills part.\footnote{More exactly, the RG equation for the effective action $\Gamma$ reads $$\mu \pd{}{\mu} \Gamma + \sum\limits_i \beta^i \partial_i \Gamma = \int d^4 x \sqrt{-g}\left( c C_{\mu\nu\rho\sigma} C^{\mu\nu\rho\sigma} - a \epsilon^{\mu\nu\alpha\beta} \epsilon^{\rho\sigma\gamma\delta} R_{\mu\nu\rho\sigma} R_{\alpha\beta\gamma\delta} \right)\,.$$ The right hand side vanishes since the Euler density $\epsilon^{\mu\nu\alpha\beta} \epsilon^{\rho\sigma\gamma\delta} R_{\mu\nu\rho\sigma} R_{\alpha\beta\gamma\delta}$ is topological and vanishes when integrated over space-time, and the Weyl tensor $C_{\mu\nu\rho\sigma} = 0$ since the cosmological backgrounds are conformally flat. Hence, if no operators except the gluon condensate are present and $\beta_{YM} \rightarrow 0$ in the IR, the theory approaches an IR fixpoint when the dilaton approaches a constant in the infrared. This is the case in the Liu-Tseytlin like backgrounds considered here, since \cite{Andrey} there the beta function vanishes in spite of the presence of a gluon condensate.}
We thus would naively expect quark confinement for $\lambda >-(4\mu^2)\sqrt{\tilde{c}_0}$
due to the strong infrared coupling. However, in this case the Wilson
loop calculation of section~\ref{confinement} shows that the quarks
are not confined: The Wilson loop deconfines due to the presence of the horizon at $g_{tt}=0$. 
We find confinement for $\lambda \leq -(4\mu^2)\sqrt{\tilde{c}_0}$ instead, where the 
coupling constant is finite and not so large, but where the horizon is absent. 
We thus conclude that similarly to the situation in the Sakai-Sugimoto
model \cite{SS}, the main factor
 controlling the confinement dynamics in this setup is not the coupling constant (i.e. the running dilaton) but the presence of a horizon in the bulk. The dilaton running is, however, important at zero temperature ($C=0$) and leads to Wilson loop confinement e.g. in the hyperbolic case $k=-1$, as will be discussed in sec.~\ref{confinement}.
}

\subsection{Boundary Energy-Momentum Tensor and Boundary Diffeomorphism Invariance}\label{4.4}

\subsubsection{The VEV of the Boundary Energy Momentum Tensor}

Next, we calculate the four-dimensional stress tensor from holography. The Fefferman-Graham expansion of the metric \eqref{finite-c-sol-3} reads 
\beqa
 \extd s^2 &=& \frac{d \rho^2}{4\rho^2} + \frac{g_{\mu\nu}(\rho,x^\mu) d x^\mu d x^\nu}{\rho}\,,\quad \mu,\nu=0,\dots,3\,,\\
 {g}_{\mu\nu}(\rho,x^\mu)&=&{g}_{(0)\mu\nu}+{g}_{(2)\mu\nu}\rho+
 {\rho^2}\left({g}_{(4)\mu\nu}+{h}_{1(4)\mu\nu}\log\rho
+{h}_{2(4)\mu\nu}(\log\rho)^2\right)+\cdots\, ,\\
 {g}_{(0)\mu\nu}&=&({g}_{(0)00},~{g}_{(0)ij})=(-1,~a_0(t)^2\gamma_{ij})\, , \quad 
 {g}_{(2)\mu\nu}=-{\lambda\over 2R^2\mu^2}{g}_{(0)\mu\nu}\, , \quad
\eeqa\label{Feff3}
and
\beq
 {g}_{(4)00}={48\tilde{c}_0-\lambda^2/\mu^2\over 16R^4}\, , \quad
 {g}_{(4)ij}={16\tilde{c}_0+\lambda^2/\mu^2\over 16R^4}{g}_{(0)ij}.
\eeq\label{Feff4}
Then by using the general formula \cite{KSS}
\beq
 \langle T_{\mu\nu}\rangle={4R^3\over 16\pi G_N}\left({g}_{(4)\mu\nu}-
 {1\over 8}{g}_{(0)\mu\nu}\left( ({\rm Tr}g_{(2)})^2-{\rm Tr}g_{(2)}^2\right)
   -{1\over 2}\left({g}_{(2)}^2\right)_{\mu\nu}+{1\over 4}{g}_{(2)\mu\nu}
    {\rm Tr}g_{(2)}\right)\, ,
\eeq\label{Feff5}
we find the holographic stress energy tensor
\beq
 \langle T_{\mu\nu}\rangle=\langle \tilde{T}_{\mu\nu}^{(0)}\rangle+
{4R^3\over 16\pi G_N^{(5)}}\left({-3\lambda^2\over 16}{g}_{(0)\mu\nu}\right)\, ,
\label{Feff6}
\eeq
\beq\label{Feff6-2}
   \langle \tilde{T}_{\mu\nu}^{(0)}\rangle={4R^3\over 16\pi G_N^{(5)}}
{\tilde{c}_0\over R^4}(3,~{g}_{(0)ij})\,,
\eeq
where $\langle \tilde{T}_{\mu\nu}^{(0)}\rangle$ is the ``thermal'' stress tensor contribution, i.e. the contribution which would be thermal for the ${\cal N}=4$ SYM fields if the universe was not expanding. The second term, which depends on $\lambda$, comes from the loop corrections of the SYM fields in a  curved
space-time, and is the conformal anomaly contribution. 
The first term does not contribute to
the conformal anomaly as in the usual finite temperature case. The conformal anomaly then reads
\beq
 \langle T_{\mu}^{\mu}\rangle=-{3\lambda^2\over 8\pi^2}N^2\, ,\label{anomaly2}
\eeq
where we used $G_N^{(5)}=8\pi^3{\alpha'}^4g_s^2/R^5$ and $R^4=4\pi N{\alpha'}^2g_s$. The conformal anomaly precisely matches the free field theory result which can be easily obtained using the general formulae given in \cite{birrell,Duff93}. 
This is expected since the conformal anomaly is one-loop exact, and hence trivially interpolates from weak to strong coupling.

\subsubsection{Time Independence of the Boundary Cosmological Constant from Diffeomorphism Invariance}\label{442}

{The stress-energy tensor \eqref{Feff6} has, in flat space ($k=0$),
  the form of a perfect fluid stress-energy tensor. As such, it obeys
  the continuity equation  
\beq
 \dot{\rho}+3H(\rho +p)=0\, ,
 \label{continuity}
\eeq
where $\rho$, $p$ and $H$ represent the energy density, pressure and the Hubble 
constant $H=\dot{a}_0/a_0$, and dot denotes the time derivative. From \eqref{Feff6} we read off
\beq
 \rho=3 \alpha \left({\tilde{c}_0\over R^4}+{\lambda^2\over 16}\right)\, ,
 \quad p=\alpha \left({\tilde{c}_0\over R^4}-3{\lambda^2\over 16}\right)\, ,
 \quad \alpha={4R^3\over 16\pi G_N^{(5)}} \label{density}\,.
\eeq
{\textit{A priori}, from the way how we introduced the boundary cosmological constant $\lambda(t)$ in sec.~\ref{sec22}, it has to be considered as  being time-dependent. The continuity equation then requires }
\beq\label{lambdaconst}
   \dot{\lambda}=0\,,
\eeq
where we used $\tilde{c}_0=C/(4\mu^2{a_0}^4(t))$. Continuity of energy density and pressure thus dictate the cosmological constant to be an actual constant in time.
}

{
The requirement that the boundary stress-energy tensor satisfies the continuity equation \eqref{continuity} is  very natural from the point of view of diffeomorphism invariance of the boundary theory. Gauge/gravity duality provides in particular a way to holographically calculate the generating functional of correlators in the dual field theory via the Gubser-Klebanov-Polyakov-Witten relation \cite{MGW} from the appropriately renormalised on-shell gravity action. Since the regularised on-shell action as well as the holographic counterterms are invariant under boundary diffeomorphisms $\xi^\mu(x^\rho)$  by construction, the resulting generating functional respects this invariance in the absence of external sources. The boundary diffeomorphisms are the ones compatible with the Fefferman-Graham expansion of the metric, and hence can only depend on boundary  coordinates $x^\rho$. In particular, only the leading piece of the expansion, the boundary metric, transforms under boundary diffeomorphisms as $g^{(0)}_{\mu\nu}\mapsto g^{(0)}_{\mu\nu} + \partial_{( \mu } \xi_{ \nu )}$, but all the subleading coefficients such as the stress-energy tensor VEV $g^{(4)}_{\mu\nu}$ will be invariant. The source-operator coupling $\int d^{p} x g^{(0)\mu\nu} T_{\mu\nu}$ then enforces the Ward identities 
\beq\label{diffeoward}
0 = \nabla^\mu \langle T_{\mu\nu} \dots \rangle 
\eeq
upon transformation of the effective action by such a boundary diffeomorphism. 
This identity should hold for correlators involving the stress-energy tensor.  In particular, the holographic stress-energy tensor should be conserved,
\beq\label{stressenergyconservation}
0 = \nabla_\mu \langle T^\mu{}_\nu \rangle\,.
\eeq
For a perfect fluid $T^\mu{}_\nu = \text{diag}(- \rho(t), p(t), p(t), p(t) )$, eq.~\eqref{stressenergyconservation} is equivalent to the continuity equation \eqref{continuity}. The time-independence of $\lambda$, eq.~\eqref{lambdaconst}, thus follows directly from the requirement of stress-energy conservation.
}

{In a more general setup one would however expect the boundary
  cosmological constant to change with time due to energy exchange
  between bulk and boundary.  As noted before, this is not possible in
  the holographic setup, due to the imposed Dirichlet boundary
  conditions imposed at the boundary, and due to the fact that gravity decouples from the gauge theory in this case as the UV cutoff is taken to infinity \cite{Gubser}. This dictates that the energy-momentum contributions to the holographic stress-energy tensor coming from the bulk (the dark radiation part) as well as from the nontrivial boundary geometry (the boundary cosmological constant part) cannot mix with each other in a nontrivial (time-dependent) way. Hence, each of them is conserved by itself, yielding \eqref{lambdaconst}.
}

\section{Dark Radiation, Boundary Cosmological Constant and Quark Confinement}\label{confinement}

{In this section we consider the combined effect of both the dark
  radiation term and the boundary cosmological constant on infinitely
  heavy quarks, i.e. test quarks, in the Super Yang-Mills theory, by holographically evaluating the static quark-antiquark potential from a Wilson loop vacuum expectation value.} 
One way to introduce (supersymmetric) quarks in the present context is through probe D7 branes \cite{D7}. 
The test quark-antiquark pair is then described by a string worldsheet ending on a prescribed space-time contour on the D7 brane. If the D7 brane corresponds to an infinitely massive embedding, and barring special issues such as the presence of gauge field charge on the brane, the brane embedding then will coincide with the asymptotic boundary of our space-time, and we can consider a string worldsheet ending on a contour at this boundary. The static quark-antiquark potential is then calculated from the energy of the string, evaluated on a minimal surface with the boundary condition prescribed by the contour \cite{WL}. Usually, the string worldsheet then has two possible configurations:
\begin{enumerate}

\item A pair of parallel strings, which stretch between the boundary and the horizon. This configuration describes a free quark-antiquark pair, and corresponds to a deconfined situation in which the Wilson loop shows a perimeter law.

\item A U-shaped string whose two end-points are on the boundary, but which does not touch any black hole horizon or singularity in the bulk. If the $g_{xx}$ component of the string frame metric has a minimum, the string will be stuck there and the energy will depend linearly (for large separations) on the separation of the end-points, and show an area law for the Wilson loop. This configuration describes a confined quark-antiquark pair. 
\end{enumerate}
These two types of configurations are seen to compete thermodynamically in the finite temperature gauge theory \cite{GSUY}, as well as for
the theory in dS$_4$ \cite{GIN1}, with the deconfined configuration being thermodynamically preferred in both cases. 

\subsection{The Wilson Loop in Cosmological Evolution}

Following \cite{WL}, we consider 
the Nambu-Goto string dynamics with the string world volume in $(t,x)$ plane. The energy $E$ 
of this state is then obtained as a function of the proper distance $L$ between 
the quark and antiquark as follows \cite{GIN1}: Choosing a gauge $X^0=t=\tau$ and $X^1=x^1=\sigma$
for the world sheet coordinates $(\tau,~\sigma)$ of the Nambu-Goto action, 
the Nambu-Goto Lagrangian in the present background (\ref{finite-c-sol-2}) 
becomes
\beq
   L_{\textrm{\scriptsize NG}}=-{1 \over 2 \pi \alpha'}\int d\sigma ~e^{\Phi/2}
   \bar{n}(r)\sqrt{r'{}^2
        +\left({r\over R}\right)^4 \left(\bar{A}(r)a_0(t)\gamma(x)\right)^2 
} ,
 \label{ng}
\eeq
where only the radial coordinate $r(x)$ is assumed to depend on $x$, and 
 prime denotes the derivative with respect to $x$. The functions $\bar n$, $\bar A$ are defined in \eqref{finite-c-sol-3}~and~\eqref{abarnbar}.

The energy of the string configuration, which is nothing but the static quark-antiquark potential, is obtained from (\ref{ng}) as
\beq
 E=-L_{\rm NG}
={1\over 2\pi \alpha'} \int d\tilde{\sigma}~ |n_s|~
        \sqrt{1+ \left({R^2\over r^2 \bar{A}}\partial_{\tilde{\sigma}}r
               \right)^2}\ , \label{W-energy}
\eeq
where 
\beq
  \tilde{\sigma}=a_0(t)\int d\sigma\gamma(\sigma)
       =a_0(t)\int d\sigma {1\over 1+k\sigma^2/4}\, ,
\eeq
and 
\beq\label{exactns}
 n_s=e^{\Phi/2}\left({r \over R}\right)^2 |\bar{A}\bar{n}|\, .
\eeq
{We should note that $\tilde{\sigma}$ measures the physical length (proper distance) in the
present case, while ${\sigma=x}$ measures the distance in comoving coordinates, which do 
not change with expanding scale. We will consider the static quark-antiquark potential as a function of proper distance.}

The quark-antiquark potential \eqref{W-energy} shows a scaling with distance if $n_s(r)$ has a finite minimum at some distance $r=r^*$ outside the horizon. This is the case for $\lambda \leq -4\mu^2\tilde{c}_0^{1/2}$. In this case 
the dilaton in $n_s(r)$ varies very slowly and monotonically (see the discussion around figure~\ref{WL-t}), and 
we can estimate the minimum of $n_s$ by neglecting the dilaton dependence, $e^{\Phi/2}\approx 1$, in $n_s(r)$, i.e. taking
\beq
n_s\approx \left({r \over R}\right)^2 |\bar{A}\bar{n}| = \left({r \over R}\right)^2 \left| \left(1-{\lambda\over 4\mu^2}\left({R\over r}\right)^2\right)^2-\tilde{c}_0 \left({R\over r}\right)^{4} \right|\,.
\eeq
We then find the minimum of $n_s(r)$ at 
\beq\label{rstar}
r^*=R\left(\left(\lambda\over 4\mu^2\right)^2-\tilde{c}_0\right)^{1/4}\,. 
\eeq
We see that the minimum is at a finite value of $r^*$ for $\lambda^2 > (4\mu^2)^2 \tilde c_0 > 0$, since $\tilde c_0>0$ is necessary in order to have a positive temperature contribution of the Yang-Mills fields to the holographic energy-momentum tensor \eqref{Feff6-2}. There are thus two regimes, $\lambda > 4\mu^2 \sqrt{\tilde c_0}$ and $\lambda <- 4\mu^2 \sqrt{\tilde c_0}$. In the former case, however, the dilaton $e^\Phi$ diverges at a finite radius, and cannot be neglected any more in \eqref{exactns}. We are thus left with considering the case $\lambda < - 4\mu^2 \sqrt{\tilde c_0}$. 
For $\lambda < - 4 \mu^2 \sqrt{\tilde c_0}$, the value of $n_s$ at the minimum is
\beq
  n_s(r^*)=
  {\lambda\over 2\mu^2} - 2\sqrt{\left({\lambda\over 4\mu^2}\right)^2-\tilde{c}_0} \geq 0\,.
\eeq
This is finite since we are considering the case of $\lambda< -4\mu^2\tilde{c}_0^{1/2}$. Note that $n_s(r^*)\geq 0$ since $\tilde c_0 \geq 0$. Then the energy 
$E$ is approximated as \cite{GIN1} 
\beq
 E\sim {n_s(r^{*})\over 2\pi \alpha'} {L}\ , \label{linear-P}
\eeq 
where 
\beq
  {L}=2\int_{\tilde{\sigma}_{min}}^{\tilde{\sigma}_{max}}d \tilde{\sigma}\, 
\eeq
is the proper distance between the string endpoints, and $\tilde{\sigma}_{min}$ ($\tilde{\sigma}_{max}$) is the value at
$r_{min}$ ($r_{max}$) of the string configuration \cite{GIN2}. 
The potential $V=E$ thus grows linearly in proper distance as long as $n_s(r^\ast)>0$, and the string tension is given by
\beq \label{tension2}
 \tau_{q\bar{q}}={n(r^{*})\over 2\pi \alpha'}\, .
\eeq
\begin{figure}[htbp]
\begin{center}
\includegraphics[width=10.0cm,height=5cm]{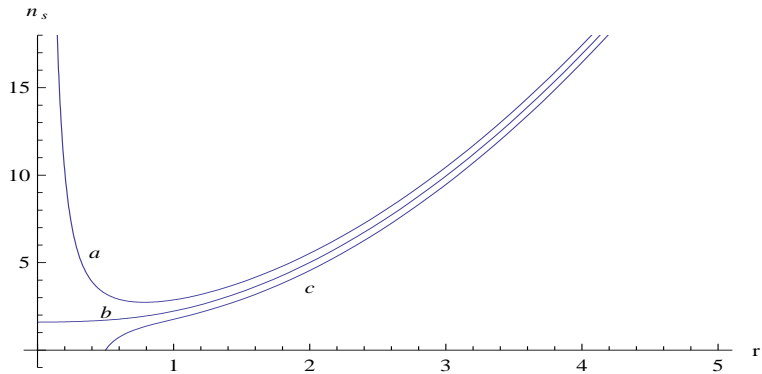}
\caption{Plots of $n$ vs $r$ for (a) $\lambda=-1-4\mu^2\tilde{c}_0^{1/2}$,
(b) $\lambda=-4\mu^2\tilde{c}_0^{1/2}$ and
(c) $\lambda=1-4\mu^2\tilde{c}_0^{1/2}$. Cases (a) and (c) are taken
as examples for $\lambda < -4\mu^2\tilde{c}_0^{1/2}$ and $\lambda >
-4\mu^2\tilde{c}_0^{1/2}$, respectively. In agreement with
\eqref{rstar}, (a) and (b) show minima and hence confine. The
parameters are taken to be $1/\mu=R=1$, $k=-1$, and
$\tilde{c}_0=0.2$. For (c), there is a horizon at $r=0.5$. Note that for the parameter values chosen in this plot, $\lambda <0$ in all cases and hence $k=-1$ is consistent when solving the Friedmann equation \eqref{Friedman}. {$k$ only enters the dilaton and hence the Wilson loop via the value of $\tilde{c}_0$. 
Furthermore, since in  general $\tilde{c}_0$ is explicitly time-dependent, these curves 
represent snapshots at constant times.}
\label{WL-t-two}}
\end{center}
\end{figure}

{On the other hand, for the case of $\lambda> -4\mu^2\tilde{c}_0^{1/2}$ we do not find such a finite minimum
of $n_s(r)$. In this case $n_s(r_H)=0$ at the horizon $r_H$ defined by $\bar n (r_H) = 0$. }
The exact behaviour of $n_s$, including the dilaton dependence, is
shown in the numerically obtained plot Fig. \ref{WL-t-two} for all
cases of relevance. The numerical results support the approximations made above. In particular, there is no linear potential for $\lambda > - 4 \mu^2 \tilde{c}_0^{1/2}$. The quarks are deconfined in this case, which can qualitatively be understood as the effect of the cosmological constant not being sufficiently negative to overcome the thermal screening of the quark-antiquark force. This is particularly interesting for negative $\lambda$, in which case we find a possible deconfined phase even for AdS backgrounds if only the finite ``temperature'' screening of the quark-antiquark potential, set in this case by the dark radiation constant $C$, is strong enough. {This seems to be a novel phenomenon at strong coupling, since the Wilson loop in AdS spaces previously was expected to confine \cite{CW,HMR} due to the diverging gravitational potential in AdS space (for a more thorough discussion see section~\ref{sec:classification}.).
}

\begin{figure}[htb]
\begin{center}
\includegraphics[width=10.0cm,height=7cm]{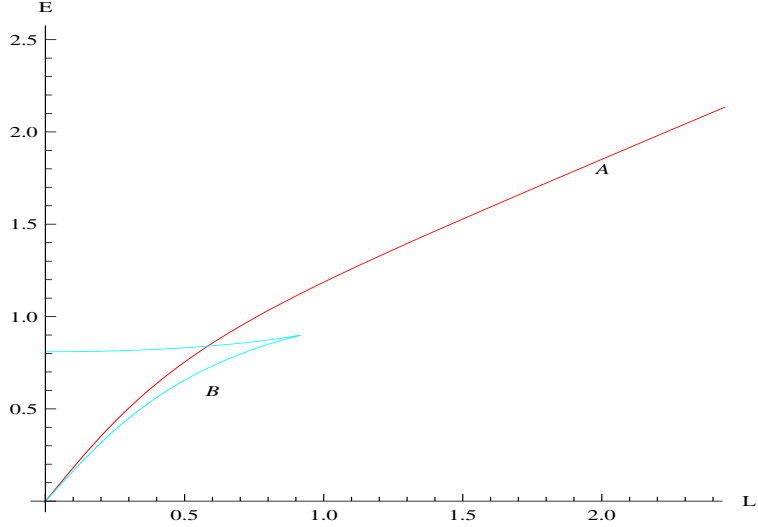}
\caption{Plots of $E$ vs ${L}$ for (A) $\lambda=-1-4\mu^2\tilde{c}_0^{1/2}$ and
(B) $\lambda=+1-4\mu^2\tilde{c}_0^{1/2}$. Cases (A) and (B) are taken as examples for $\lambda < -4\mu^2\tilde{c}_0^{1/2}$ and $\lambda > -4\mu^2\tilde{c}_0^{1/2}$, respectively. Case (A) shows deconfinement, while case (B) shows typical screening behaviour: In the latter case there exist two U-shaped configurations for small $L$, while for large separations the string breaks at the horizon and hence has vanishing energy (at $O(N^2)$). 
Other parameters are set as $q=0$, $\mu=1$, $R=1$, $k=-1$, 
$\tilde{c}_0=1.0$,
$r_{\textrm{\scriptsize max}}=3\, $ 
and $\alpha'=1\,$. {Note that in general $\tilde{c}_0$ is explicitly time-dependent, these curves 
represent snapshots of the dilaton solution at constant time.}
\label{el}}
\end{center}
\end{figure}

We obtained the relation $E(L)$, as shown in fig.~\ref{el}, and
the tension $\tau_{q\bar{q}}$ 
in the following way: Since the Lagrangian in (\ref{ng}) does not explicitly depend on 
the coordinate $\sigma=x$, we find the following quantity conserved under $\sigma$-shifts,
\beq
     e^{\Phi/2}{1\over \sqrt{(r/R)^4 \bar{A}^2(r)+(r')^2}}
    \left({r\over R}\right)^4 \bar{n}\bar{A}^2(r)= H\,.
\eeq
We can fix $H$ at any point we like, so we fix it at $r=r_{min}$. Then, choosing 
$H=e^{\Phi/2}\left({r\over R}\right)^2 \bar{n}(r)\bar{A}(r)|_{r_{min}}$, we get
\bea
  &&  {L}=2R^2 \int_{r_{min}}^{r_{\textrm{\scriptsize max}}} dr~
      {1\over r^2 \bar{A}(r)
        \sqrt{e^{\Phi(r)}r^4 \bar{n}(r)^2\bar{A}(r)^2 /
          \left(e^{\Phi(r_{min})}r_{min}^4
       \bar{n}(r_{min})^2\bar{A}(r_{min})^2\right)-1}} , \nonumber %\label{len}
\\
  && E=
   {1\over \pi \alpha'} \int_{r_{min}}^{r_{\textrm{\scriptsize max}}}dr~
   {\bar{n}(r)e^{\Phi(r)/2}\over 
     \sqrt{1-e^{\Phi(r_{min})}r_{min}^4 \bar{n}(r_{min})^2\bar{A}(r_{min})^2/
             \left(e^{\Phi(r)}r^4 \bar{n}(r)^2\bar{A}(r)^2\right)}} . \label{energy}
\eea
Figure~\ref{el} shows the exact dependence of the energy $E$ 
on the distance ${L}$ for the values $q=0$ (i.e. for constant dilaton)
for $\lambda\leq -4\mu^2\tilde{c}_0^{1/2}$ 
(curve A) and $\lambda >-4\mu^2\tilde{c}_0^{1/2}$ (curve B). 
In the former case, we find 
the linear potential at large $L$ as expected, and we find a 
typical screening behavior 
in the latter case, similar to the one seen in the finite temperature deconfinement phase. 
The qualitative behaviour of the Wilson loop for $q\neq 0$ are unchanged from the $q=0$ case. 

\subsection{Classification of (De)confining Behaviour in Cosmological Backgrounds}\label{sec:classification}

{
In the above analysis we found confining behaviour for the Wilson loop if 
\beq\label{confiningWL}
\lambda \leq - 4 \mu^2 \sqrt{\tilde c_0} = - \frac{2 \sqrt{C/R^2}}{a_0^2(t)}\,.
\eeq
{Thus for vanishing dark radiation constant $C=0$ (which would correspond to vanishing temperature in the static case) and negative boundary cosmological constant $\lambda<0$ the system is always confined, in accordance with earlier results of \cite{CW,AMR}.} 
On the other hand, the Wilson loop cannot have an area law for
\beq\label{deconfiningWL}
\lambda > - 4 \mu^2 \sqrt{\tilde c_0} = - \frac{2 \sqrt{C/R^2}}{a_0^2(t)}\,.
\eeq
This regime is deconfining. The right-hand side of these inequalities is generically time-dependent, and hence the inequalities will hold only in particular time intervals, in which the universe is larger/smaller than a critical value set by the boundary cosmological constant. There are three cases:
\begin{enumerate}

\item Positive $\lambda$: In this case the system is in a deconfined phase. This could have been expected by the expanding nature of the de Sitter universe in this case: The expansion of space tends to destabilise bound states, leading to deconfinement even at zero dark radiation constant.

\item Vanishing $\lambda $: This case is similar to the (de)confinement properties of a planar AdS-Schwarzschild black hole. 
For any finite value of the dark radiation constant $C>0$ (``finite temperature''), eq.~\eqref{deconfiningWL} is satisfied, and the system is in the deconfined phase. 
For vanishing dark radiation constant $C=0$, however, eq.~\eqref{confiningWL} holds and the Wilson loop shows confining behaviour.

\item The most interesting situation occurs for negative
  $\lambda$: Depending on the value of the scale factor $a_0(t)$
  at any given time, either \eqref{confiningWL} or
  \eqref{deconfiningWL} can be satisfied, leading to 
  transitions between confined and deconfined phases as the universe evolves. {Physically,
    this transition is due to the competition between two effects: The
    screening effects of the thermal (dark radiation) energy always
    aims at driving the system into a deconfined phase, while the influence of the background cosmological evolution on the static quark-antiquark potential can be either confining (for $\lambda < 0$) or deconfining (for $\lambda > 0$). Thus for negative boundary cosmological constant thermal screening and background-induced confinement can compete and result into (de)confinement transitions.}

For definiteness, let us consider the solution of \eqref{Friedman} for $k=-1$ and $\lambda<0$,
\beq\label{atsolklambdaneg}
a_0(t) = \frac{\sin \sqrt{|\lambda|} t}{\sqrt{|\lambda|}}\,.
\eeq
This solution thus describes an oscillating universe. 
Analysing \eqref{confiningWL}, \eqref{deconfiningWL}, we find two different regimes: Since the scale factor is bounded from above by $a_{0,max}=1/\sqrt{|\lambda|}$, a large enough dark radiation constant $C > R^2/4$ always satisfies \eqref{deconfiningWL}, and hence quarks are always deconfined in this case. For $C < R^2/4$, however, the system oscillates between a deconfined phase at smaller  scale factors $a_0(t)$, and a confined phase at larger values. For the marginal value $C = R^2/4$, the Wilson loop shows confining behaviour only at maximal extension of the universe. Our holographic setup thus describes a (supersymmetric) plasma with the qualitative  properties observed in the evolution of our universe: Near the big bang singularity $a_0=0$ the matter is in a deconfined state, and undergoes a confinement phase transition as the universe cools down. Figure~\ref{deconftransition} summarises the situation. 

These results, in particular in the latter case, need to be compared to the
results of \cite{HMR}, where it was argued that the Wilson loop is not a
good measure for (de)confinement in AdS space. 
The arguments of \cite{HMR} involve a conformal transformation between AdS
space and half of the Einstein static universe (ESU), 
 relating long distance behaviour in AdS space to the 
(universal) short distance behaviour in the ESU measured in turn by
the Wilson loop. This argument fails in the backgrounds considered
here since the conformal  symmetry of ${\cal N}=4$ SYM theory is
broken by the gluon condensate, as well as by the conformal anomaly for $\lambda  \neq 0$.\footnote{This point was noted before in \cite{Buchel}.}  The results 
of sec.~\ref{confinement} show that the Wilson loop is sensitive to both, 
and hence depends on the chosen conformal frame, measuring unambiguously the deconfinement properties of the 
chosen field theory state by coupling to the full energy-momentum
tensor and to the gluon condensate.\footnote{The gluon condensate 
is essential for confinement in the case $\lambda=0$, $k=-1$, since the Wilson loop in the purely hyperbolic AdS-Schwarzschild black hole is screened \cite{LL} for all temperatures.}   
Similarly, we do not expect the arguments of \cite{CW} for a weakly coupled 
meron gas disordering the Wilson to simply carry over to strong coupling.

The conformal anomaly is also responsible for the time-dependent nature of the holographic backgrounds presented here: 
The diffeomorphism relating these backgrounds to topological black holes \cite{Tetradis1,Tetradis2} induces a conformal transformation on the boundary, which due to the conformal anomaly is not a quantum symmetry of the dual field theory. Thus, except for the case of vanishing boundary cosmological constant,  observables calculated in the two different conformal frames will generically be different, and therefore have to be calculated in the time-dependent background itself. 
We cannot resort to the equivalence with topological black holes
to define a thermodynamic ensemble or to consider phase transitions,
as e.g. studied for  compactifications of CFT's on dS
space-times  \cite{dSLiterature}. Our setup hence is time-dependent and
 describes genuine non-equilibrium  physics. 
We can however characterize the properties of the non-equilibrium state by calculating observables such as the Wilson loop via the machinery of gauge-gravity duality.

For vanishing boundary cosmological constant the conformal anomaly
vanishes, and the phase structure of the AdS-Schwarzschild black hole
with flat ($k=0$) horizon, which has $O(N^2)$ entropy density at
finite temperature and $O(1)$ at absolute zero \cite{WittenThermal}, 
coincides exactly with our results for the Wilson loop. 
For the hyperbolic ($k=-1$) black hole, on the other hand, these two
measures of confinement do not agree: From \cite{Tetradis1,Tetradis2}
it is clear that the black hole mass is given by $\mu = C$. The
hyperbolic black hole has a nondegenerate horizon at $r_+=L_{AdS}$ for
$\mu=0$, and  both its free energy and entropy are $O(N^2)$ at this
point \cite{Myersetal}. Our Wilson loop on the other hand is confined
at $\mu=C=0$. This is one of the rare examples where the density of
states and the Wilson loop do not agree as measures of confinement,
which can be traced back in this case to the effect of the gluon
condensate which enters the Wilson loop via the string frame metric.

As discussed in the introduction, different measures of confinement
leading to different answers in curved backgrounds are not uncommon,
and since the field theory is known exactly,\footnote{The field theory
 is ${\cal
    N}=4$ SYM theory with a gluon condensate \cite{LT}. This
breaks conformal symmetry
  spontaneously, but not explicitly
  at the  level of symmetry generators or Green functions, since the
  beta function still vanishes (see also \cite{Andrey}).} this case appears to be
a good playground for a future investigation of the interplay of
different measures of confinement.\footnote{This is particularly
  interesting in view of claims that the so-called `precursor' states
  \cite{hyperbolic}, which create the $O(N^2)$ ground state entropy of
  the extremal hyperbolic black hole, are potentially relevant to the thermal
  screening of the quark-antiquark potential in the absence of the
  gluon condensate \cite{LL}.} We plan to come back to this
point  in a future work.

Another interesting observation concerns the relation of the Wilson loop
with the temperature \eqref{temperature2}: Although this temperature
has been derived in a adiabatic approximation, assuming the cosmological constant to be sufficiently small, the Wilson loop feels exactly this temperature, without any approximation. The reason is that the Wilson loop calculation in this section is done ``locally in time'', i.e. by considering a string stretching into the fifth dimension at each  fixed value in time, testing the presence of the horizon with \eqref{temperature2}. If the horizon is present the Wilson loop  exhibits perimeter law, if not, the temperature \eqref{temperature2} is zero or ill-defined, and the Wilson loop exhibits an area law. Thus, although \eqref{temperature2} can only be considered as an approximation, it exactly reproduces the Wilson loop behaviour.

\end{enumerate}

%%%%%%%%%%%%%% Fig %%%%%%%%%%%%%%%%%%%%% ->
\begin{figure}[htb]%[H]
\begin{center}
\includegraphics[width=10.0cm]{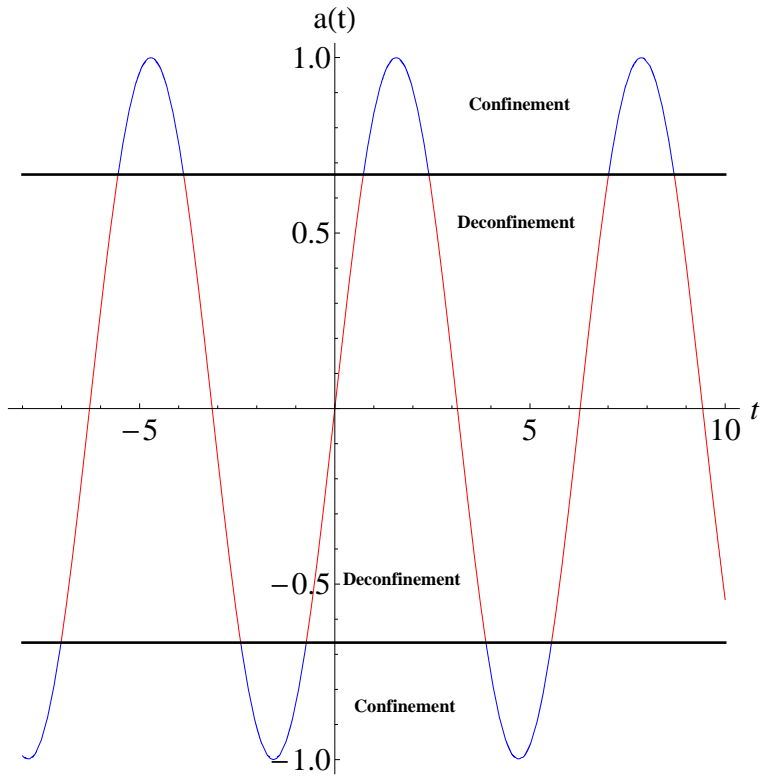}
\caption{The (de)confinement transition as seen in the Wilson loop behaviour in an oscillating, open, universe ($\lambda <0$, $k=-1$).  During the periods depicted in blue, the Wilson loop shows confining behaviour, while during the red periods (close to the big bang singularity), the plasma is in a deconfined phase. The transition line, shown in black, is fixed by the value of the dark radiation constant $C$.
\label{deconftransition}}
\end{center}
\end{figure}
%%%%%%%%%%%%%%%%%%%%%%%%%%%%%%%%%%%%%%%%% <-

}

\section{Summary and Discussions}\label{summary}

{

We have investigated properties of strongly coupled 
${\cal N}=4$ $U(N)$ supersymmetric Yang-Mills theory in the presence of a gluon condensate on cosmological space-times 
of Friedmann-Robertson-Walker type by applying methods of gauge/gravity duality. 
The dual gravity solutions are obtained from a Liu-Tseytlin Ansatz  
of IIB supergravity by solving the effective five-dimensional Einstein equations 
through a metric Ansatz first employed in the setup of braneworld cosmologies 
\cite{BDEL}. %The Liu-Tseytlin Ansatz in particular allows the study of field theory dynamics in the presence of gluon condensates and instanton densities. 
By introducing a single integration constant, the so-called ``dark radiation constant''
 or ``mirage energy density'', Einstein's equations reduce to a single constraint 
 equation. We solved this equation by introducing a boundary cosmological constant which \textit{a priori} can be time-dependent.  In this way the constraint is separated into a standard Friedmann equation and an equation for the sole undetermined function in the bulk metric. The resulting holographic background is dual to ${\cal N}=4$ SYM theory in a FRW-type metric, 
 and has a the gluon condensate. This bulk geometry determines the vacuum expectation value of the stress tensor. 

Holographically, the dark radiation term \cite{Kiritsis:2005bm,Tetradis1,Tetradis2} 
 induces a relativistic radiation component (of Stefan-Boltzmann form) in the stress-energy tensor. If the universe is static, this radiation component has the correct $T^4$ behaviour and 
$N^2$ scaling in the large $N$ limit. If the universe is nonstatic, it is modified by a factor $1/a_0^4(t)$, as
expected for a gas of relativistic particles. For nonvanishing boundary cosmological constant there is also the familiar conformal anomaly contribution proportional to the square of the boundary cosmological constant $\lambda^2$.

Using the holographic stress-energy tensor, we have also clarified the holographic 
interpretation of a possible time-dependence of the cosmological constant $\lambda(t)$: The holographic stress-energy tensor is conserved if and only if the boundary cosmological constant $\lambda(t)$ is time independent, $\dot \lambda = 0$. Requiring a conserved boundary stress-energy tensor is necessary to ensure that the dual field theory is boundary diffeomorphism invariant. In general relativist's terms, the coupling between field theory and curved background geometry does not spoil the equivalence principle. Hence, with the standard AdS/CFT Dirichlet boundary conditions, only a time-independent boundary cosmological constant $\lambda(t)=\lambda$ is holographically meaningful and consistent.

The main result of this paper is the behaviour of the Wilson loop in
the cosmological background geometries, which we take as the measure of the
(de)confinement properties of quark-antiquark pairs. We find 
an interesting interplay between dark radiation and cosmological
constant: The dark radiation component drives the Wilson loop to
deconfining behaviour, which may be understood as thermal screening of the quark-antiquark interaction. On the other hand, the boundary cosmological constant $\lambda$ can have deconfining or confining effect: Positive $\lambda$ (i.e. de Sitter like expanding cosmologies) drives the system into deconfinement, while negative $\lambda$ (i.e. anti-de Sitter like cosmologies) drives it towards confinement. For negative $\lambda$ there exist periodically oscillating cosmologies in which the system undergoes periodic (de)confinement transitions: For dark radiation constants below a critical value set by the bulk AdS radius, the Wilson loop is deconfined when the universe is small (i.e. near the big bang singularity). After sufficient expansion it undergoes a transition to confining behaviour.  This is in qualitative agreement with the expected behaviour in nature: Close to the big bang, i.e. at large temperatures, QCD matter should have been in a quark-gluon plasma state, and undergoes a confinement phase transition once the universe sufficiently expanded and cooled down.

It should be stressed that in contrast to previous works \cite{Kiritsis:2005bm,Tetradis1,Tetradis2}, the standard holographic Dirichlet boundary conditions  employed in this paper forbid bulk-boundary energy exchange. The dual field theory lives on a curved but fixed background geometry with no propagating graviton in the boundary theory. The Friedmann equation is obtained from the bulk Einstein equations. It is then natural that boundary diffeomorphism invariance restricts the choice of bulk geometry.

As a follow-up and 
in view of the interesting (de)confinement properties of the system 
described, it appears to be worthwhile 
to investigate the dynamics of fundamental degrees of freedom. Such
 degrees of freedom can e.g. be introduced into these cosmological backgrounds by probe D7 branes \cite{D7}. This would for instance allow an investigation of the chiral symmetry breaking and its relation to Wilson loop (de)confinement, as well as possible chiral symmetry enhancement or suppression. -- 
Applying holographic renormalisation to these backgrounds will further clarify the relation between the free energy, entropy, and the properties of the Wilson loop, and thus shed light on the relation between these different measures of confinement. In a similar way the existence of a mass gap can be inferred from a fluctuation analysis around the bulk geometry. 
We will come back to these questions in a  future work.
}
                                                                 
\section*{Acknowledgments}

We are indebted to Elias Kiritsis for discussions on the results of
this paper, to Mukund Rangamani to correspondence and discussions on
\cite{HMR,AMR}, to Curtis Callan for a useful correspondence on 
\cite{CW}, to Matt Lippert, Takeshi Morita and George Moutsopoulos for  discussions and to Masafumi Ishihara for collaboration at early stages of this work. The research of R.~M. is supported 
by the European Union grant FP7-REGPOT-2008-1-CreteHEPCosmo-228644. R.~M. would thank the Fukuoka Institute of Technology, CQUeST \& Sogang U. (Seoul), the Erwin-Schr\"odinger-Institute (Vienna), the Galileo Galilei Institute (Florence), the KITPC and the ITP-CAS (Beijing), and the Max Planck Institut f\"ur Physik (M\"unchen) for hospitality while this research was in progress.

%%%%%%%%%%%%%%%%%%%%%%%
%%%%%%%%%%%%%%%%  References %%%%%%%%%%%%%%%%%%

\newpage

\begin{thebibliography}{99}

\bibitem{MGW}
J.~M.~Maldacena,
%``The large N limit of superconformal field theories and supergravity,''
Adv.\ Theor.\ Math.\ Phys.\  {\bf 2}, 231 (1998) [hep-th/9711200].
%%CITATION = HEP-TH 9711200;%%

S.~S.~Gubser, I.~R.~Klebanov and A.~M.~Polyakov,
%``Gauge theory correlators from non-critical string theory,''
Phys.\ Lett.\ B {\bf 428}, 105 (1998) [hep-th/9802109].
%%CITATION = HEP-TH 9802109;%%

E.~Witten,
%``Anti-de Sitter space and holography,''
Adv.\ Theor.\ Math.\ Phys.\  {\bf 2}, 253 (1998) [hep-th/9802150].
%%CITATION = HEP-TH 9802150;%%

A.~M.~Polyakov,
  %``The wall of the cave,''
  Int.\ J.\ Mod.\ Phys.\  A {\bf 14}, 645 (1999)
  [arXiv:hep-th/9809057].
  %%CITATION = IMPAE,A14,645;%%

\bibitem{Gubser:1999pk}
  S.~S.~Gubser,
  %``Dilaton-driven confinement,''
  arXiv:hep-th/9902155.
  %%CITATION = HEP-TH/9902155;%%
\bibitem{KS2}
  A.~Kehagias and K.~Sfetsos,
  %``On asymptotic freedom and confinement from type-IIB supergravity,''
  Phys.\ Lett.\  B {\bf 456}, 22 (1999)
  [arXiv:hep-th/9903109].
  %%CITATION = PHLTA,B456,22;%%
\bibitem{LT}
  H.~Liu and A.~A.~Tseytlin,
  %``D3-brane D-instanton configuration and N = 4 super YM theory in  constant
  %self-dual background,''
  Nucl.\ Phys.\  B {\bf 553} (1999) 231
  [arXiv:hep-th/9903091].
  %%CITATION = NUPHA,B553,231;%%

%%%%%% AdS/dS
\bibitem{Hawking}
  S.~Hawking, J.~M.~Maldacena and A.~Strominger,
  %``DeSitter entropy, quantum entanglement and AdS/CFT,''
  JHEP {\bf 0105}, 001 (2001)
  [arXiv:hep-th/0002145].
  %%CITATION = JHEPA,0105,001;%%

%%%%%% dS/dS
\bibitem{Alishahiha1}
  M.~Alishahiha, A.~Karch, E.~Silverstein and D.~Tong,
  %``The dS/dS correspondence,''
  AIP Conf.\ Proc.\  {\bf 743}, 393 (2005)
  [arXiv:hep-th/0407125].
  %%CITATION = APCPC,743,393;%%
\bibitem{Alishahiha2}
  M.~Alishahiha, A.~Karch and E.~Silverstein,
  %``Hologravity,''
  JHEP {\bf 0506} (2005) 028
  [arXiv:hep-th/0504056].
  %%CITATION = JHEPA,0506,028;%%
  
  
%%%%%%AdS/dS                                                                           
\bibitem{H}
  T.~Hirayama,
  %``A holographic dual of CFT with flavor on de Sitter space,''
  JHEP {\bf 0606}, 013 (2006)
  [arXiv:hep-th/0602258].
  %%CITATION = JHEPA,0606,013;%%

\bibitem{GIN1}
  K.~Ghoroku, M.~Ishihara and A.~Nakamura,
  %``Gauge theory in de Sitter space-time from a holographic model,''
  Phys.\ Rev.\  D {\bf 74}, 124020 (2006)
  [arXiv:hep-th/0609152].
  %%CITATION = PHRVA,D74,124020;%%

%%%%%%AdS/AdS 
\bibitem{GIN2}
  K.~Ghoroku, M.~Ishihara and A.~Nakamura,
  %``Flavor quarks in AdS(4) and gauge/gravity correspondence,''
  Phys.\ Rev.\  D {\bf 75}, 046005 (2007)
  [arXiv:hep-th/0612244].
  %%CITATION = PHRVA,D75,046005;%%

\bibitem{GSUY} 
  K.~Ghoroku, T.~Sakaguchi, N.~Uekusa and M.~Yahiro,
  %``Flavor quark at high temperature from a holographic model,''
  Phys.\ Rev.\  D {\bf 71}, 106002 (2005)
  [arXiv:hep-th/0502088].
  %%CITATION = PHRVA,D71,106002;%%

\bibitem{AMR}
O.~Aharony, D.~Marolf and M.~Rangamani,
  %``Conformal field theories in anti-de Sitter space,''
  arXiv:1011.6144 [hep-th].
  %%CITATION = ARXIV:1011.6144;%%

\bibitem{Kraus:1999it}
  P.~Kraus,
  %``Dynamics of anti-de Sitter domain walls,''
  JHEP {\bf 9912} (1999) 011
  [arXiv:hep-th/9910149].
  %%CITATION = JHEPA,9912,011;%%

\bibitem{Kehagias:1999vr}
  A.~Kehagias and E.~Kiritsis,
  %``Mirage cosmology,''
  JHEP {\bf 9911} (1999) 022
  [arXiv:hep-th/9910174].
  %%CITATION = JHEPA,9911,022;%%

\bibitem{BDEL} 
  P.~Binetruy, C.~Deffayet, U.~Ellwanger and D.~Langlois,
  %``Brane cosmological evolution in a bulk with cosmological constant,''
  Phys.\ Lett.\  B {\bf 477}, 285 (2000)
  [arXiv:hep-th/9910219].
  %%CITATION = PHLTA,B477,285;%%

\bibitem{Lang} 
  D.~Langlois,
  %``Brane cosmological perturbations,''
  Phys.\ Rev.\  D {\bf 62}, 126012 (2000)
  [arXiv:hep-th/0005025].
  %%CITATION = PHRVA,D62,126012;%%
  D.~Langlois and L.~Sorbo,
  %``Bulk gravitons from a cosmological brane,''
  Phys.\ Rev.\  D {\bf 68}, 084006 (2003)
  [arXiv:hep-th/0306281].
  %%CITATION = PHRVA,D68,084006;%%

\bibitem{Tetradis1}
 P.~S.~Apostolopoulos, G.~Siopsis and N.~Tetradis,
  %``Cosmology from an AdS Schwarzschild black hole via holography,''
  Phys.\ Rev.\ Lett.\  {\bf 102} (2009) 151301
  [arXiv:0809.3505 [hep-th]].
  %%CITATION = PRLTA,102,151301;%%

\bibitem{Tetradis2}
  N.~Tetradis,
  %``The temperature and entropy of CFT on time-dependent backgrounds,''
  JHEP {\bf 1003} (2010) 040
  [arXiv:0905.2763 [hep-th]].
  %%CITATION = JHEPA,1003,040;%%

\bibitem{LL}
  K.~Landsteiner and E.~Lopez,
  %``Probing the strong coupling limit of large N SYM on curved backgrounds,''
  JHEP {\bf 9909} (1999) 006
  [arXiv:hep-th/9908010].
  %%CITATION = JHEPA,9909,006;%%

\bibitem{Kiritsis:2005bm}
  E.~Kiritsis,
  %``Holography and brane-bulk energy exchange,''
  JCAP {\bf 0510} (2005) 014
  [arXiv:hep-th/0504219].
  %%CITATION = JCAPA,0510,014;%%

\bibitem{Brito}
  L.~Barosi, F.~A.~Brito and A.~R.~Queiroz,
  %``Holographic Description of Heavy-Quark Potentials in an Inflationary
  %Braneworld Scenario,''
  JHEP {\bf 0904} (2009) 030
  [arXiv:0812.4841 [hep-th]].
  %%CITATION = JHEPA,0904,030;%%

%%%%%%%% Field Theory in AdS stuff
\bibitem{CW}
  C.~G.~Callan and F.~Wilczek,
  %``INFRARED BEHAVIOR AT NEGATIVE CURVATURE,''
  Nucl.\ Phys.\  B {\bf 340} (1990) 366.
  %%CITATION = NUPHA,B340,366;%%



\bibitem{HMR}
  V.~E.~Hubeny, D.~Marolf and M.~Rangamani,
  %``Hawking radiation from AdS black holes,''
  Class.\ Quant.\ Grav.\  {\bf 27} (2010) 095018
  [arXiv:0911.4144 [hep-th]].
  %%CITATION = CQGRD,27,095018;%%



  
\bibitem{GGP} 
  G.~W.~Gibbons, M.~B.~Green and M.~J.~Perry,
  %``Instantons and Seven-Branes in Type IIB Superstring Theory,''
  Phys.\ Lett.\  B {\bf 370}, 37 (1996)
  [arXiv:hep-th/9511080].
  %%CITATION = PHLTA,B370,37;%%

\bibitem{Marolf}
  G.~Comp\`ere and D.~Marolf,
  %``Setting the boundary free in AdS/CFT,''
  Class.\ Quant.\ Grav.\  {\bf 25} (2008) 195014
  [arXiv:0805.1902 [hep-th]].
  %%CITATION = CQGRD,25,195014;%%

%%%%%%%%%%%%%5
\bibitem{RS1} 
  L.~Randall and R.~Sundrum,
  %``A large mass hierarchy from a small extra dimension,''
  Phys.\ Rev.\ Lett.\  {\bf 83}, 3370 (1999)
  [arXiv:hep-ph/9905221].
  %%CITATION = PRLTA,83,3370;%%
\bibitem{bre} 
  I.~H.~Brevik, K.~Ghoroku, S.~D.~Odintsov and M.~Yahiro,
  %``Localization of gravity on brane embedded in AdS(5) and dS(5),''
  Phys.\ Rev.\  D {\bf 66}, 064016 (2002)
  [arXiv:hep-th/0204066].
  %%CITATION = PHRVA,D66,064016;%%
\bibitem{GY} 
  K.~Ghoroku and M.~Yahiro,
  %``Chiral symmetry breaking driven by dilaton,''
  Phys.\ Lett.\  B {\bf 604}, 235 (2004)
  [arXiv:hep-th/0408040].
  %%CITATION = PHLTA,B604,235;%%

%%%%%%%% FeffermanGraham
\bibitem{KSS} 
  S.~de Haro, S.~N.~Solodukhin and K.~Skenderis,
  %``Holographic reconstruction of spacetime and renormalization in the  AdS/CFT
  %correspondence,''
  Commun.\ Math.\ Phys.\  {\bf 217}, 595 (2001)
  [arXiv:hep-th/0002230].
  %%CITATION = CMPHA,217,595;%%
\bibitem{BFS} 
  M.~Bianchi, D.~Z.~Freedman and K.~Skenderis,
  %``Holographic Renormalization,''
  Nucl.\ Phys.\  B {\bf 631}, 159 (2002)
  [arXiv:hep-th/0112119].
  %%CITATION = NUPHA,B631,159;%%
\bibitem{FG}
{C. Fefferman and C. Robin Graham, `Conformal Invariants', in 
{\it Elie Cartan et les Math\'ematiques d'aujourd'hui} 
(Ast\'erisque, 1985) 95.}

%%%%%%%%%%%%%%%%%%%%%%%%%%%%%
\bibitem{GKP} 
  S.~S.~Gubser, I.~R.~Klebanov and A.~W.~Peet,
  %``Entropy and Temperature of Black 3-Branes,''
  Phys.\ Rev.\  D {\bf 54}, 3915 (1996)
  [arXiv:hep-th/9602135].
  %%CITATION = PHRVA,D54,3915;%%
\bibitem{NS} 
  S.~Nakamura and S.~J.~Sin,
  %``A holographic dual of hydrodynamics,''
  JHEP {\bf 0609}, 020 (2006)
  [arXiv:hep-th/0607123].
  %%CITATION = JHEPA,0609,020;%%

% AdS/CFT and Gravity
\bibitem{Gubser}
  S.~S.~Gubser,
  %``AdS/CFT and gravity,''
  Phys.\ Rev.\  D {\bf 63} (2001) 084017
  [arXiv:hep-th/9912001].
  %%CITATION = PHRVA,D63,084017;%%

\bibitem{LiPang}
  M.~Li and Y.~Pang,
  %``Holographic de Sitter Universe,''
  arXiv:1105.0038 [hep-th].
  %%CITATION = ARXIV:1105.0038;%%

\bibitem{SS}
 T.~Sakai and S.~Sugimoto,
  %``Low energy hadron physics in holographic QCD,''
  Prog.\ Theor.\ Phys.\  {\bf 113} (2005) 843
  [arXiv:hep-th/0412141], 
  %%CITATION = PTPKA,113,843;%%
  %``More on a holographic dual of QCD,''
  Prog.\ Theor.\ Phys.\  {\bf 114} (2005) 1083
  [arXiv:hep-th/0507073].
  %%CITATION = PTPKA,114,1083;%%


\bibitem{Andrey}
 J.~Erdmenger, A.~Gorsky, P.~N.~Kopnin, A.~Krikun and A.~V.~Zayakin,
  %``Low-Energy Theorems from Holography,''
  JHEP {\bf 1103} (2011) 044
  [arXiv:1101.1586 [hep-th]].
  %%CITATION = JHEPA,1103,044;%%





\bibitem{birrell} 
  N.~D.~Birrell and P.~C.~W.~Davies,
  %``Quantum Fields In Curved Space,''
%\href{/spires/find/hep/www?irn=998621}{SPIRES entry}
{\it  Cambridge, Uk: Univ. Pr. ( 1982) 340p}
\bibitem{Duff93} 
  M.~J.~Duff,
  %``Twenty years of the Weyl anomaly,''
  Class.\ Quant.\ Grav.\  {\bf 11}, 1387 (1994)
  [arXiv:hep-th/9308075].
  %%CITATION = CQGRD,11,1387;%%


\bibitem{D7}
  A.~Karch and E.~Katz,
  %``Adding flavor to AdS/CFT,''
  JHEP {\bf 0206}, 043 (2002)
  [arXiv:hep-th/0205236].\\
  %%CITATION = JHEPA,0206,043;%%
  M.~Kruczenski, D.~Mateos, R.~C.~Myers and D.~J.~Winters,
  %``Meson spectroscopy in AdS/CFT with flavour,''
  JHEP {\bf 0307}, 049 (2003)
  [arXiv:hep-th/0304032].\\
  %%CITATION = JHEPA,0307,049;%%
  M.~Kruczenski, D.~Mateos, R.~C.~Myers and D.~J.~Winters,
  %``Towards a holographic dual of large-N(c) QCD,''
  JHEP {\bf 0405}, 041 (2004)
  [arXiv:hep-th/0311270].\\
  %%CITATION = JHEPA,0405,041;%%
  J.~Babington, J.~Erdmenger, N.~J.~Evans, Z.~Guralnik and I.~Kirsch,
  %``Chiral symmetry breaking and pions in non-supersymmetric gauge /  gravity
  %duals,''
  Phys.\ Rev.\  D {\bf 69}, 066007 (2004)
  [arXiv:hep-th/0306018].
  %%CITATION = PHRVA,D69,066007;%%


\bibitem{WL}
  J.~M.~Maldacena,
  %``Wilson loops in large N field theories,''
  Phys.\ Rev.\ Lett.\  {\bf 80} (1998) 4859
  [arXiv:hep-th/9803002].\\
  %%CITATION = PRLTA,80,4859;%%
S.~J.~Rey and J.~T.~Yee,
  %``Macroscopic strings as heavy quarks in large N gauge theory and  anti-de
  %Sitter supergravity,''
  Eur.\ Phys.\ J.\  C {\bf 22}, 379 (2001)
  [arXiv:hep-th/9803001].\\
  %%CITATION = EPHJA,C22,379;%%
S.~J.~Rey, S.~Theisen and J.~T.~Yee,
  %``Wilson-Polyakov loop at finite temperature in large N gauge theory and
  %anti-de Sitter supergravity,''
  Nucl.\ Phys.\  B {\bf 527}, 171 (1998)
  [arXiv:hep-th/9803135].\\
  %%CITATION = NUPHA,B527,171;%%
 A.~Brandhuber, N.~Itzhaki, J.~Sonnenschein and S.~Yankielowicz,
  %``Wilson loops, confinement, and phase transitions in large N gauge  theories
  %from supergravity,''
  JHEP {\bf 9806}, 001 (1998)
  [arXiv:hep-th/9803263].
  %%CITATION = JHEPA,9806,001;%%

\bibitem{Buchel}
  A.~Buchel,
  %``Gauge theories on hyperbolic spaces and dual wormhole instabilities,''
  Phys.\ Rev.\  D {\bf 70} (2004) 066004
  [arXiv:hep-th/0402174].
  %%CITATION = PHRVA,D70,066004;%%

\bibitem{dSLiterature}
  O.~Aharony, M.~Fabinger, G.~T.~Horowitz and E.~Silverstein,
  %``Clean time dependent string backgrounds from bubble baths,''
  JHEP {\bf 0207}, 007 (2002)
  [arXiv:hep-th/0204158].\\
  %%CITATION = JHEPA,0207,007;%%
   V.~Balasubramanian and S.~F.~Ross,
  %``The Dual of nothing,''
  Phys.\ Rev.\  D {\bf 66}, 086002 (2002)
  [arXiv:hep-th/0205290].\\
  %%CITATION = PHRVA,D66,086002;%%
  R.~G.~Cai,
  %``Constant curvature black hole and dual field theory,''
  Phys.\ Lett.\  B {\bf 544}, 176 (2002)
  [arXiv:hep-th/0206223].\\
  %%CITATION = PHLTA,B544,176;%%
  S.~F.~Ross and G.~Titchener,
  %``Time-dependent spacetimes in AdS/CFT: Bubble and black hole,''
  JHEP {\bf 0502}, 021 (2005)
  [arXiv:hep-th/0411128].\\
  %%CITATION = JHEPA,0502,021;%%
  V.~Balasubramanian, K.~Larjo and J.~Simon,
  %``Much ado about nothing,''
  Class.\ Quant.\ Grav.\  {\bf 22}, 4149 (2005)
  [arXiv:hep-th/0502111].\\
  %%CITATION = CQGRD,22,4149;%%
  J.~He and M.~Rozali,
  %``On bubbles of nothing in AdS/CFT,''
  JHEP {\bf 0709}, 089 (2007)
  [arXiv:hep-th/0703220].\\
  %%CITATION = JHEPA,0709,089;%%
  J.~A.~Hutasoit, S.~Prem Kumar and J.~Rafferty,
  %``Real time response on dS(3): The Topological AdS Black Hole and the
  %Bubble,''
  JHEP {\bf 0904}, 063 (2009)
  [arXiv:0902.1658 [hep-th]].\\
  %%CITATION = JHEPA,0904,063;%%
  D.~Marolf, M.~Rangamani and M.~Van Raamsdonk,
  %``Holographic models of de Sitter QFTs,''
  arXiv:1007.3996 [hep-th]. 
  %%CITATION = ARXIV:1007.3996;%%
  J.~Blackman, M.~B.~McDermott and M.~Van Raamsdonk,
  %``Acceleration-Induced Deconfinement Transitions in de Sitter Spacetime,''
  arXiv:1105.0440 [hep-th].
  %%CITATION = ARXIV:1105.0440;%%

\bibitem{WittenThermal}
    E.~Witten,
  %``Anti-de Sitter space, thermal phase transition, and confinement in gauge
  %theories,''
  Adv.\ Theor.\ Math.\ Phys.\  {\bf 2}, 505 (1998)
  [arXiv:hep-th/9803131].
  %%CITATION = 00203,2,505;%%

%% Supergravity solutions
\bibitem{Myersetal}
  R.~Emparan, C.~V.~Johnson and R.~C.~Myers,
  %``Surface terms as counterterms in the AdS / CFT correspondence,''
  Phys.\ Rev.\  D {\bf 60} (1999) 104001
  [arXiv:hep-th/9903238].
  %%CITATION = PHRVA,D60,104001;%%

\bibitem{hyperbolic}
R.~Emparan,
  %``AdS / CFT duals of topological black holes and the entropy of zero energy
  %states,''
  JHEP {\bf 9906} (1999) 036
  [arXiv:hep-th/9906040].
  %%CITATION = JHEPA,9906,036;%%
G.~Horowitz, A.~Lawrence and E.~Silverstein,
  %``Insightful D-branes,''
  JHEP {\bf 0907} (2009) 057
  [arXiv:0904.3922 [hep-th]].
  %%CITATION = JHEPA,0907,057;%%








\end{thebibliography}
\end{document}